\documentclass[copyright,creativecommons]{eptcs}

\providecommand{\event}{LTT 2026} 

\usepackage{iftex}

\ifpdf
  \usepackage{underscore}         
  \usepackage[T1]{fontenc}        
\else
  \usepackage{breakurl}           
\fi

\usepackage[usenames,dvipsnames,svgnames,table]{xcolor}
\usepackage{amsmath}
\usepackage{amssymb}
\usepackage{amsthm}
\usepackage{hyperref}
\usepackage{cleveref}
\usepackage{cite}
\usepackage{wrapfig}

\usepackage{tikz}
\usetikzlibrary{arrows,shapes.geometric,shapes.misc,positioning,automata}
\usetikzlibrary{trees,decorations.pathmorphing,calc}
\usetikzlibrary{intersections}

\theoremstyle{definition}
\newtheorem{definition}{Definition}[section]

\newtheorem{example}[definition]{Example}

\usepackage{amsmath}
\usepackage{amssymb}
\usepackage{amsthm}
\usepackage{listings}
\usepackage{stmaryrd}
\usepackage[]{todonotes}
\usepackage{mathtools}
\usepackage{soul}
\usepackage{comment}

\usepackage{xspace}

\makeatletter
\newcommand\footnoteref[1]{\protected@xdef\@thefnmark{\ref{#1}}\@footnotemark}
\makeatother

\newcommand{\corrstart}{\color{black}}
\newcommand{\corrend}{\color{black}}
\newcommand{\correction}[1]{\corrstart{}#1\corrend{}}

\newcommand{\extstart}{}
\newcommand{\extension}[1]{\extstart{}#1\corrend{}}

\newcommand{\FORGET}[1]{}

\definecolor{qedgray}{rgb}{0.31,0.31,0.33}

\newcommand{\ap}[1]{{\langle #1 \rangle}}
\newcommand{\bp}[1]{{\left\lbrace #1 \right\rbrace}}

\newcommand{\auxcolour}{\color{violet}} 
\newcommand{\kcolour}{\color{blue}} 

\newcommand{\FC}{FC}

\newcommand{\name}{\tau}
\newcommand{\BNFcce}{{\textbf{::=}}}
\newcommand{\BNFmid}{\;\bigr\rvert\;}

\newcommand{\e}{\mathtt{e}}

\newcommand{\main}{\mathtt{main}}
\newcommand{\emain}{\e_{\main}}

\newcommand{\aname}{\mathtt{a}}
\newcommand{\bname}{\mathtt{b}}
\newcommand{\cname}{\mathtt{c}}

\newcommand{\oname}{\mathtt{o}}
\newcommand{\xname}{\mathtt{x}}
\newcommand{\yname}{\mathtt{y}}

\newcommand{\nname}{\mathtt{n}}
\newcommand{\sname}{\mathtt{s}}

\newcommand{\anyvalue}{\mathtt{v}}
\newcommand{\anyvaluealt}{\mathtt{w}}

\newcommand{\lvalue}{\ell}

\newcommand{\funvalue}{\mathtt{f}}

\newcommand{\FVname}{\mathsf{FV}}
\newcommand{\FV}[1]{\FVname(#1)}
\newcommand{\FTVname}{\mathsf{TV}}
\newcommand{\FTV}[1]{\FTVname(#1)}

\newcommand{\tsinsrel}{\prec}

\newcommand{\funKname}{{\kcolour\mathtt{fun}}^\name}
\newcommand{\funK}{{\kcolour\mathtt{fun}\;}}
\newcommand{\defK}{{\kcolour\mathtt{def}\;}}
\newcommand{\oldK}{{\mathtt{old}}}
\newcommand{\nbrK}{{\mathtt{nbr}}}

\newcommand{\exchangeK}{{\auxcolour\mathtt{exchange}}}
\newcommand{\returnK}{{\kcolour\mathtt{return}}}
\newcommand{\sendK}{{\kcolour\mathtt{send}}}
\newcommand{\retsendK}{{\kcolour\mathtt{retsend}}}
\newcommand{\pExchange}[4]{\exchangeK(#1, (#2) \, \toSymFoK \, \pairK(#3, #4))}
\newcommand{\dExchange}[4]{\exchangeK(#1, (#2) \, \toSymFoK \, \returnK\; #3 \; \sendK\; #4)}
\newcommand{\sExchange}[3]{\exchangeK(#1, (#2) \, \toSymFoK \, \retsendK\, #3)}
\newcommand{\vExchange}[3]{\exchangeK(#1, (#2) \, \toSymFoK \, (\valK \xname = #3;\; \pairK(\xname, \xname))}
\newcommand{\ifK}{{\kcolour\mathtt{if}}}
\newcommand{\muxK}{{\auxcolour\mathtt{mux}}}

\newcommand{\elseK}{{\kcolour\,\mathtt{else}\,}}

\newcommand{\foldK}{{\auxcolour\mathtt{nfold}}}

\newcommand{\valK}{{\kcolour\mathtt{val}\;}}

\newcommand{\uidK}{{\auxcolour\mathtt{uid}}}
\newcommand{\updateSelfK}{{\auxcolour\mathtt{updateSelf}}}
\newcommand{\updateDefK}{{\auxcolour\mathtt{updateDef}}}
\newcommand{\nbrDistK}{{\auxcolour\mathtt{senseDist}}}

\newcommand{\orK}{{\auxcolour\mathtt{or}}}
\newcommand{\andK}{{\auxcolour\mathtt{and}}}

\newcommand{\toSymFoK}{\mathrm{\texttt{=>}}}
\newcommand{\toSymK}[1][]{\stackrel{#1}{\mathrm{\texttt{=>}}}}

\newcommand{\selfK}{{\auxcolour\mathtt{self}}}

\newcommand{\type}{\textit{T}}

\newcommand{\ltype}{\textit{A}}
\newcommand{\ltypealt}{\textit{B}}

\newcommand{\ltypefun}{\textit{K}}

\newcommand{\typeOf}[2]{#1[#2]}
\newcommand{\ltypeOf}[1]{\typeOf{\ltypefun}{#1}}

\newcommand{\tbool}{\mathtt{bool}}
\newcommand{\tnum}{\mathtt{num}}

\newcommand{\tfieldOf}[1]{\underbar{\ensuremath{#1}}}
\newcommand{\ptnpair}{\mathtt{PAIR}}
\newcommand{\tpairOf}[2]{\ptnpair[#1,#2]}
\newcommand{\ptnlist}{\mathtt{LIST}}

\newcommand{\PairK}{{\mathtt{Pair}}}

\newcommand{\pairK}{{\auxcolour\mathtt{pair}}}
\newcommand{\fstK}{{\auxcolour\mathtt{fst}}}
\newcommand{\sndK}{{\auxcolour\mathtt{snd}}}

\newcommand{\tvar}{\alpha}
\newcommand{\typescheme}{\textit{TS}}

\newcommand{\TtypEnv}{\mathcal{A}}

\newcommand{\expTypJud}[3]{#1 \vdash #2 : #3}
\newcommand{\surfaceTyping}[3]{
  \begin{array}{@{\!\,}l@{\;}c@{\!\,}}
    \stackrel{~}{{\tiny \textrm{[#1]}}} & #2 \\ \hline
    \multicolumn{2}{c}{#3}
  \end{array}
}
\newcommand{\nullsurfaceTyping}[2]{
  \surfaceTyping{#1}{}{#2}
}

\newcommand{\deviceS}{\Delta}

\newcommand{\TreeSet}{\Omega}
\newcommand{\Trees}{\Theta}
\newcommand{\emptyseq}{\bullet}

\newcommand{\senstate}{\sigma}

\newcommand{\envmap}[2]{#1\mapsto #2}

\newcommand{\proj}[2]{{#1} \, |_{#2}}
\newcommand{\evalue}[3]{#1[\envmap{#2}{#3}]}

\newcommand{\ruleNameSize}[1]{{\scriptsize #1}}

\newcommand{\vtree}{\theta}
\newcommand{\mkvt}[2]{#1 \langle #2 \rangle}

\newcommand{\piI}[1]{\pi_{#1}}
\newcommand{\piIof}[2]{\piI{#1}(#2)}

\newcommand{\bsopsem}[6]{#1;#3;#2\vdash #4\Downarrow #5; #6}
\newcommand{\eqhlauxopsem}[6]{#1;#3;#2\vdash #4\Downarrow^\ast #5; #6}
\newcommand{\eqhlbsopsem}[6]{#1;#3;#2\vdash #4\Downarrow #5; #6}

\newcommand{\deviceId}{\delta}

\newcommand{\substitution}[2]{#1:=#2}
\newcommand{\applySubstitution}[2]{#1[#2]}
\newcommand{\nameOf}{\textit{name}}

\newcommand{\skiptransition}{\\~\\[-10pt]}

\newcommand{\aEventS}[0]{\mathbb{E}}

\newcommand{\eventS}[0]{E}
\newcommand{\eventId}[0]{\epsilon}

\newcommand{\setVS}[0]{\mathbf{V}}

\newcommand{\neigh}{\rightsquigarrow}

\newcommand{\devof}{d}
\newcommand{\sensof}{s}

\newcommand{\dvalue}[0]{\mathrm{\Phi}}

\definecolor{dark-gray}{gray}{0}

\renewcommand{\ldots}{%
\mathinner{\ldotp\mkern-4mu\ldotp\mkern-4mu\ldotp}%
}

\lstdefinelanguage{afc}{
	basicstyle=\small\ttfamily\lst@ifdisplaystyle\scriptsize\fi,
	frame=single,
	basewidth=0.50em,
	sensitive=true,
	morestring=[b]",
	morecomment=[l]{//},
	morecomment=[n]{/*}{*/},
	commentstyle=\color{OliveGreen},
	keywordstyle=\color{blue}\textbf, keywords={def,if,else,val,return,send,retsend,type,place},  otherkeywords={@},
	keywordstyle=[3]\color{violet}, keywords=[3]{exchange,nfold,self,updateSelf,updateDef,uid,pow,min,and,or,mux,distance,pair,fst,snd,head,tail},
	keywordstyle=[4]\color{darkgray}\textbf, keywords=[4]{D,E,T,U,Num,Bool,Pair,List,
            Task,Point,Msg,File,Option,True,False,Infinity,None,Some,DB},
	keywordstyle=[5]\color{RoyalPurple}, keywords=[5]{rep,nbr,share,broadcast,serve_data,current_task,towards_avoiding,scout_target},
}

\lstdefinelanguage{fcpp}{
	basicstyle=\footnotesize\ttfamily\lst@ifdisplaystyle\footnotesize\fi,
	frame=single,
	basewidth=0.5em,
	sensitive=true,
	morestring=[b]",
	morecomment=[l]{//},
	morecomment=[n]{/*}{*/},
	commentstyle=\color{OliveGreen},
	keywordstyle=\color{blue}, keywords={return,auto,FUN,GEN,ARGS,CODE,CALL},
	keywordstyle=[2]\color{violet}\textbf, keywords=[2]{bool,int,double,field},
	keywordstyle=[3]\color{orange}, keywords=[3]{fold_hood,min_hood,mux,min,max,INF,or,and},
	keywordstyle=[4]\color{red}, keywords=[4]{exchange,nbr,rep,share}
}

\newcommand{\sysname}{{\textsf{XC}}\xspace}

\newcommand{\excvalue}[0]{nvalue}
\newcommand{\excvalues}{\excvalue{}s}
\newcommand{\Excvalue}[0]{Nvalue}
\newcommand{\Excvalues}{\Excvalue{}s}

\definecolor{darkgreen}{rgb}{0,0.5,0}

\lstdefinelanguage{scafi}{
  basicstyle=\normalsize\ttfamily\lst@ifdisplaystyle\footnotesize\fi,
	frame=single,
	escapechar=\%,
  keywords={abstract,case,catch,class,def,%
    do,else,extends,final,finally,%
    for,if,implicit,import,match,mixin,%
    new,null,object,override,package,%
    private,protected,requires,return,sealed,%
    super,this,throw,trait,try,lazy,%
    type,val,var,while,with,yield,forSome},
  otherkeywords={=>,<-,<\%,<:,>:,\#},
  keywordstyle=\color{red}\textbf,
  keywordstyle=[2]\color{blue},
  keywords=[2]{exchange,branch,@@},
  keywordstyle=[3]\color{violet},
  keywords=[3]{fold,mux,collect,pair,broadcast,foldSum,distanceTo,channel,timestamp,channelBroadcast},
  keywordstyle=[4]\color{Emerald},
  keywordstyle=[5]\color{Brown},
  sensitive=true,
  morecomment=[l]{//},
  morecomment=[n]{/*}{*/},
  commentstyle=\color{darkgreen},
  morestring=[b]",
  morestring=[b]',
  morestring=[b]"""
}

\newcommand{\supplier}[0]{supplier}
\newcommand{\XC}{XC}

\newcommand{\temperatureK}{{\auxcolour\mathtt{temperature}}}
\newcommand{\gpsK}{{\auxcolour\mathtt{gps}}}
\newcommand{\timeK}{{\auxcolour\mathtt{time}}}

\title{Lambdas at the Far Edge: a Tale of\\ Flying Lambdas and Lambdas on Wheels}

\author{Giorgio Audrito
$\;\;$
Daniele Bortoluzzi
$\;\;$
Ferruccio Damiani
$\;\;$
Giordano Scarso
$\;\;$
Gianluca Torta
\institute{
Department of Computer Science (DI) 
\\
University of Turin, Turin, Italy}
\email{name.surname@unito.it}
\and
Andrea Basso
\institute{MITO Technology, Milan, Italy}
\email{speedrun2006@gmail.com}
\and
Monica Cochi 
$\;\;$
Lorenzo Gusman
\institute{
Torino Airport\\
Caselle Torinese (TO), Italy}
\email{name.surname@sagat.trn.it}
\and
Lorenzo Comba 
$\;\;$
Paolo Gay  
\institute{
Department of Agricultural, Forest and Food Sciences (DiSAFA)
\\
University of Turin, Turin, Italy}
\email{name.surname@unito.it}
\and
Paola Dal Zovo 
\institute{
Concept Engineering\\
Reply, Turin, Italy} 
\email{p.dalzovo@reply.it}
\and
Giada Galati 
$\;\;$
Francesco Gallo 
\institute{
Eurix, Turin, Italy}
\email{surname@eurixgroup.com}
\and
Aljaž Grdadolnik
\institute{Faculty of Computer and Information Science\\
University of Ljubljana, Ljubljana, Slovenia}
\email{aljaz.grdadolnik1@gmail.com}
\and
Massimo Pescarollo
$\;\;$
Paola Pisano 
\institute{
Department of Economics and Statistics ``Cognetti de Martiis''\\
University of Turin, Turin, Italy}
\email{name.surname@unito.it}
}

\def\titlerunning{Lambdas at the Far Edge}
\def\authorrunning{G. Audrito et al.}
\begin{document}

\lstset{language=afc}

\maketitle

\begin{abstract}
Aggregate Programming (AP) is a paradigm for programming the collective behaviour of sets of distributed devices, possibly situated at the network far edge, by relying on asynchronous proximity-based interactions.
The eXchange Calculus (XC), a recently proposed foundational model for AP, is essentially a typed lambda calculus extended with an operator (the exchange operator) providing an implicit communication mechanism between neighbour devices. This paper provides a gentle introduction to XC and to its implementation as a C++ library, called FCPP. The FCPP library and toolchain has been mainly developed at the Department of Computer Science of the University of Turin, where Stefano Berardi spent most of his academic career conducting outstanding research about logical foundation of computer science and transmitting his passion for research to students and young researchers, often exploiting typed lambda calculi. An FCCP program is essentially a typed lambda term, and FCPP has been used to write code that has been deployed on devices at the far edge of the network, including rovers and (soon) Uncrewed Aerial Vehicles (UAVs); hence the title of the paper. 
\end{abstract}


\begin{flushright}
    \footnotesize To Stefano Berardi on the occasion of his $2^6$th birthday.
\end{flushright}

\section{Introduction} \label{sec:introduction}

Embedded systems, smart dust, ubiquitous/pervasive computing, Cyber-Physical Systems and Internet of Things are buzzwords
(listed roughly in chronological order)
used by researchers and practitioners to indicate sensing, computing and actuation entities
(called nodes, devices, etc.)
deployed over a physical environment. A clear trend of these technologies is the increase in number of deployed entities, making it infeasible to deploy a global-level software functionality by individually programming every single entity.
For this reason, a series of approaches -- which can be classified under "macroprogramming" and "spatial self-organization" (surveyed in
\cite{SpatialIGI2013,autonomicommunications,DBLP:journals/jsa/MameiMTZ06}) --  are investigated. They are all aimed at providing means to view a network of nodes as a single spatially distributed platform 
which could be programmed as a whole.

Among these approaches, the \emph{Aggregate Programming (AP)}~\cite{BPV-COMPUTER2015,DBLP:journals/jlap/ViroliBDACP19} paradigm is based on understanding system interactions in terms of information that propagates through collectives of devices, possibly (but not necessarily~\cite{DBLP:conf/isola/AudritoDT22}) situated at the extreme edge of the network, relying on implicit interactions based on proximity, rather than
 explicit interactions of individual devices with their peers and with the environment.
The reference programming models for AP are the \emph{Field Calculus (FC)}~\cite{DVB-SCP2016,Viroli:HFC-TOCL,DBLP:journals/corr/abs-1910-02874} and -- its recently proposed generalization -- the \emph{eXchange Calculus (XC)}~\cite{DBLP:conf/ecoop/AudritoCDSV22,Audrito-et-al:JSS-2024,Audrito-et-al:COORDINATION-2024}. They are core functional languages enriched with proximity-based interaction primitives which allow programmers to structure complex computational tasks by bottom-up construction of reusable blocks of intrinsically resilient aggregate behaviour. In particular, XC can be understood as a typed lambda calculus enriched with an operator (the exchange operator) providing an implicit communication mechanism between neighbour devices.
 
The main practical support for AP are the C++ library and toolchain \emph{FCPP} \cite{FCPP-ACSOS-2020,DBLP:conf/coordination/AudritoRT22,DBLP:journals/tpds/AudritoTF23,DBLP:journals/scp/AudritoT24} (mainly developed 
at the Department of Computer Science of the University of Turin) 
and the Scala library and toolchain \emph{ScaFi} \cite{Casadei:PMLDC16,DBLP:journals/softx/CasadeiVAP22,DBLP:conf/acsos/CasadeiAPV23,DBLP:journals/lmcs/AudritoCDV23}
(mainly developed at the Department of Computer Science and Engineering of the University of Bologna).
FCPP allows programmers to embed XC programs in C++ applications, and deploy them on networks of physical devices, as well as a simulator.  
 It is used in a number of industrial case studies currently developed in collaboration with the companies Reply (\url{https://www.reply.com}), Synesthesia (\url{https://synesthesia.it}), Eurix (\url{https://www.eurixgroup.com/})
 and Torino Airport (\url{https://www.aeroportoditorino.it/en}).\footnote{See, e.g., the RoboNG (\url{https://www.ecs-nodes.eu/en/1-aerospace-and-sustainable-mobility/progetti-imprese/robong}) and the RoboAPP  (\url{https://www.ecs-nodes.eu/en/1-aerospace-and-sustainable-mobility/progetti-imprese/robonghttps://ecs-nodes.eu/en/1-aerospace-and-sustainable-mobility/progetti-accademici/roboapp}) 
 cascade funding projects of the Spoke 1 ``Sustainable mobility and aerospace'' of the  PNRR NODES (\url{https://ecs-nodes.eu/en}); and the CN AgriTech (\url{https://agritechcenter.it/}).}
 -- a preliminary result of these activities was recently published~\cite{DBLP:journals/percom/TestaADT22}. 

\paragraph*{Contribution of the paper}
In this paper, we provide a gentle introduction to XC tailored for researchers who, like Stefano Berardi, have research interests in lambda calculus and type theory. We also briefly introduce the FCPP library and toolchain and briefly review some industrial case studies where FCCP programs, which can be understood as typed lambda terms, have been deployed on rovers and are going to be deployed on Unmanned Aerial Vehicles (UAVs) -- hence the title of the paper. Stefano Berardi spent most of his academic career at the Department of Computer Science of the University of Turin, where he conducts outstanding research on the logical foundations of computer science and passes on his passion for research to students and young researchers (including -- in reverse chronological order -- the first author and the third author of this paper), often leveraging typed lambda calculus. We therefore believe that the research activities reported in this paper fully fall within Stefano Berardi's research interests.

\paragraph*{Organization of the paper}
\Cref{sec:fromLCtoXC} provides a formal account (syntax, typing and operational semantics) of XC, while 
\Cref{sec:design-and-properties} briefly outlines XC design decisions and main properties.
\Cref{sec:fcpp} briefly introduces the FCPP library and toolchain.
\Cref{sec:fcpp-at-work} briefly overviews some industrial case studies.
\correction{
\Cref{sec:relwork} discusses related work.
}
\Cref{sec:conclusion} concludes with final remarks and directions of future work.

\section{From the Lambda Calculus to the eXchange Calculus} \label{sec:fromLCtoXC}

\subsection{The Lambda Calculus inside the eXchange Calculus} \label{sec:the-lambda-calculus-inside-xc}


In this section we present the syntax and typing of \XC, which can be understood as a call-by-value lambda calculus typed \`a la ML.

\subsubsection{Syntax} \label{ssec:pfc:syntax}

\begin{figure*}[tb]
\centerline{\framebox[\linewidth]{$\begin{array}{l}
	\textbf{Syntax:} \\
	\begin{array}{c@{\hspace{3pt}}c@{\hspace{3pt}}l@{\hspace{21mm}}r}
		\e & \BNFcce &
									\xname
				\, \BNFmid \, \funK \xname(\overline\xname) \{ \e \}
				\, \BNFmid \, \e(\overline\e)
				\, \BNFmid \, \valK \; \xname = \e_1; \e_2
				\, \BNFmid \, \lvalue
		& {\footnotesize \mbox{expression}}
		\\[3pt]
		\lvalue & \BNFcce & \bname \, \BNFmid \, \funK \xname(\overline\xname) \{ \e \} \, \BNFmid \, \cname(\overline\lvalue)
		& {\footnotesize \mbox{literal value}}
		\\[3pt]
		\bname & \BNFcce &  \ldots
		& {\footnotesize \mbox{built-in function}}
	\end{array} \\
		\hline\\[-10pt]
	\textbf{Free variables of an expression:} \hspace{59mm} \boxed{\FV{\e}}
	\\
	\begin{array}{ll}
		\FV{\xname} = \{ \xname \}  \quad \FV{\lvalue}  = \emptyset
		&
		\FV{\funK \xname_0(\xname_1,\ldots,\xname_n) \{ \e \}} = \FV{\e} \setminus \bp{\xname_0,\ldots,\xname_n}
		\\[2pt]
		\medmuskip=1mu
		\thinmuskip=1mu
		\thickmuskip=1mu
		\nulldelimiterspace=0pt
		\scriptspace=0pt
		\FV{\e_0(\e_1,\ldots,\e_n)} \,=\, \bigcup_{i=0 \ldots n} \FV{\e_i}
		&
		\FV{\valK \; \xname = \e_1; \e_2} = \FV{\e_1} \cup (\FV{\e_2} \setminus \bp{\xname})
	\end{array}\\
	\hline\\[-10pt]
	\textbf{Syntactic sugar:} \\
	\begin{array}{c@{\hspace{3pt}}c@{\hspace{3pt}}l@{\hspace{-9mm}}r}
		(\overline\xname) \toSymK \e & \BNFcce & \funK \yname(\overline\xname) \{ \e \} \textit{ where } \yname \textit{ is a fresh variable}
  & {\footnotesize \mbox{lambda expression}}
		\\[3pt]
		\defK \xname(\overline\xname) \{ \e \} & \BNFcce & \valK \xname = \funK \xname(\overline\xname) \{ \e \};
  & {\footnotesize \mbox{top-level function definition}}
		\\[3pt]
		\ifK (\e) \{\e_1\} \elseK \{\e_2\} & \BNFcce & \muxK(\e, () \toSymK \e_1, () \toSymK \e_2)()
  & {\footnotesize \mbox{conditional expression}}
	\end{array}
\end{array}$}}
\caption{\XC: syntax (top), free variables (middle) and   syntactic sugar (bottom).} \label{fig:pfc:syntax}
\end{figure*}

The syntax of \XC\ is given in \Cref{fig:pfc:syntax} (top).
Following~\cite{FJ}, the overbar notation indicates a (possibly empty) sequence of elements, e.g., $\overline{x}$ is short for $x_1, \ldots, x_n$ $(n\ge 0)$. 
An \XC\ \emph{expression} $\e$ can be either:
\begin{itemize}
	\item
	a \emph{variable} $\xname$;
	\item
	a (possibly recursive)  \emph{function} $\funK \xname(\overline\xname) \{ \e \}$, which may have free variables;
	\item
	a \emph{function call} $\e(\overline\e)$;
	\item
	a \emph{let-like} expression $\valK \xname = \e; \e$;
	\item
	a \emph{literal value} $\lvalue$, that is either a built-in function $\bname$, a defined function $\funK \xname(\overline\xname) \{ \e \}$ \emph{without} free variables, or a data constructor $\cname$ applied to literal values (possibly none).
\end{itemize}
Built-ins -- listed in \Cref{fig:standard} -- include the multiplexed function $\muxK$, an eager conditional operator which evaluates all its three arguments and returns the second or the third according to the value of the first; and built-in functions and data constructors for standard data types like numbers, Booleans, pairs, lists, etc.

\begin{figure*}[tb]
\begin{scriptsize}
\centerline{
\begin{tabular}{@{\hspace{7pt}}c@{\hspace{7pt}}|@{\hspace{7pt}}c@{\hspace{7pt}}|@{\hspace{7pt}}l@{\hspace{7pt}}}
\textsc{Name} & \textsc{Type scheme} & \textsc{Description}
\\ \hline
\multicolumn{3}{l}{\textbf{Functions:}}
\\\hline
$\muxK$ & $(\tbool,\ltype,\ltype) \to \ltype$  & \emph{Multiplexer operator}
\\ \hline
\lstinline|+|, \lstinline|-|, \lstinline|*|, \lstinline|/| &
$(\tnum,\tnum) \to \tnum$ & \emph{Arithmetic operators}
\\\hline
$\andK$, $\orK$ &
$(\tbool,\tbool) \to \tbool$ & \emph{Boolean operators}
\\ \hline
$\mathtt{==}$, $\mathtt{<=}$, $\mathtt{>=}$ & $(\ltype,\ltype) \to \tbool$ & \emph{Relational operators}
\\ \hline
$\pairK$ & $(\ltype,\ltypealt) \to \tpairOf{\ltype}{\ltypealt}$ & \emph{Pair creation}
\\ \hline
$\fstK$ & $(\tpairOf{\ltype}{\ltypealt}) \to \ltype$ & \emph{First element of a pair}
\\ \hline
$\sndK$ & $(\tpairOf{\ltype}{\ltypealt}) \to \ltypealt$ & \emph{Second element of a pair}
\\ \hline
\multicolumn{3}{l}{\textbf{Data constructors:}}
\\\hline
$\mathtt{-1}$, $\mathtt{0}$, $\mathtt{0.25}$, $\mathtt{1}$, $\mathtt{Infinity}$ & $\tnum$ & \emph{Numeric constructors} \\ \hline
$\mathtt{True}$, $\mathtt{False}$ & $\tbool$ & \emph{Boolean constructors}
\\ \hline
$\PairK$ & $(\ltype,\ltypealt) \to \tpairOf{\ltype}{\ltypealt}$ & \emph{Pair constructor}
\\ \hline
\end{tabular}
}
\end{scriptsize}
\caption{\XC: name, type scheme and description of  built-in functions and data constructors used in the examples -- the generic $\ltype$ type in relational operators is not allowed to contain occurrences of a function type.} \label{fig:standard}
\end{figure*}

Free variables are defined  in a standard way (\Cref{fig:pfc:syntax}, middle), and an expression $\e$ is \emph{closed} if $\FV{\e}=\emptyset$. Programs are closed expressions. 

 Besides some standard syntactic sugar (infix notation for binary operators, omitted parenthesis in 0-ary data constructors), some non-standard syntactic sugar is described in \Cref{fig:pfc:syntax} (bottom). In particular: a lambda expression  is encoded as a fun-expression with a fresh name \correction{(i.e., not appearing anywhere else in the whole program)};
a top-level function definition is encoded as a let-like expression binding the function name; and a conditional expression
is encoded by lambda-abstracting the code in the branches, selecting one of them with the $\muxK$ operator and then applying it.


\subsubsection{Typing} \label{ssec:typing}

\Cref{fig:pfc:typing} presents a classic Hindley-Milner type system~\cite{Damas-Milner:POPL-1982} for \XC.
 A type $\type$ can be:
\begin{itemize}
	\item
	a \emph{type variable} $\tvar$;
	\item
	a (possibly recursive) \emph{data type} $\ltypeOf{\overline\type}$, consisting of a \emph{parametric type name}  $\ltypefun$ (like, e.g., $\tnum$, $\tbool$, $\ptnpair$, $\ptnlist$,...) of arity $n\ge 0$, applied to $n$ types $\type_1,...,\type_n$ (possibly zero) -- as syntactic sugar, we write $\ltypefun$ as short for $\ltypefun[]$ (e.g., $\tnum$
 as short for $\tnum[]$);
	\item
	or a \emph{function type} $(\overline\type) \rightarrow \type$.
\end{itemize}
We write $\FTV{\type}$ for the set of type variables in $\type$. Polymorphic functions and data  constructors are supported by type schemes $\typescheme$ of the form $\forall \overline{\tvar}.\type$ where $\overline{\tvar}$ occur free in ${\type}$, representing all types obtained by substituting $\overline{\tvar}$ with types $\overline{\type}$, as per the type scheme instantiation relation $\tsinsrel$.
\begin{figure*}[tb]
\centerline{\framebox[\linewidth]{$\begin{array}{l}
	\textbf{Types, type schemes, typing environments  and instantiation:} \\
	\begin{array}{c@{\hspace{3pt}}c@{\hspace{3pt}}l@{\hspace{-7mm}}r @{\hspace{5mm}} c@{\hspace{3pt}}c@{\hspace{3pt}}l@{\hspace{4mm}}r}
		\type & \BNFcce & \tvar \, \BNFmid \, \ltypeOf{\overline\type} \, \BNFmid \, (\overline\type) \rightarrow \type
		& {\footnotesize \mbox{type}}
		&
		\aname & \BNFcce & \xname \, \BNFmid \, \bname \, \BNFmid \,  \cname
		& {\footnotesize \mbox{assumption subject}}
		\\[4pt]
		\typescheme & \BNFcce & \forall \overline{\tvar}.\type
		& {\footnotesize \mbox{type scheme}}
		&
		\TtypEnv & \BNFcce & \correction{\overline\aname : \overline{\typescheme}}
		& {\footnotesize \mbox{typing environment}}
		\\[4pt]
		\forall\overline{\tvar}.\type 	 & \tsinsrel  & \type\,[\overline{\tvar}:=\overline{\type}]   & {\footnotesize
		\mbox{instantiation}}
	\end{array} \\
	\hline\\[-10pt]
	\textbf{Expression typing:} \hspace{75mm} \boxed{\expTypJud{\TtypEnv}{\e}{\type}} \\
	\begin{array}{c}
		\surfaceTyping{T-ASS}{ ~
			\TtypEnv(\aname) \tsinsrel \type \text{ for } \aname = \xname \text{ or } \bname
		}{
			\expTypJud{\TtypEnv}{\aname}{\type}
		}
		\qquad
			\surfaceTyping{T-LIT}{ ~
			\TtypEnv(\cname) \tsinsrel (\overline\type)\rightarrow\type \quad
			\expTypJud{\TtypEnv}{\overline\lvalue}{\overline{\type}}
		}{
			\expTypJud{\TtypEnv}{\cname(\overline\lvalue)}{\type}
		}
		\skiptransition
		\surfaceTyping{T-APP}{ ~
			\expTypJud{\TtypEnv}{\e}{(\overline\type)\rightarrow\type} \qquad
			\expTypJud{\TtypEnv}{\overline\e}{\overline{\type}}
		}{
			\expTypJud{\TtypEnv}{\e(\overline\e)}{\type}
		}
		\skiptransition
		\surfaceTyping{T-FUN}{ ~
			\expTypJud{\TtypEnv, \xname : \forall \emptyseq. (\overline\type) \to \type, \overline\xname : \forall \emptyseq. \overline\type}{\e}{\type} \qquad
		}{
			\expTypJud{\TtypEnv}{\funK \xname(\overline\xname)\{\e\}}{(\overline\type) \to \type}
		}
		\skiptransition
		\surfaceTyping{T-VAL}{ ~
			\expTypJud{\TtypEnv}{\e_1}{\type_1} \qquad
			\overline\tvar = \FTV{\type_1} \qquad
			\expTypJud{\TtypEnv,\xname : \forall \overline{\tvar}.\type_1}{\e_2}{\type_2} \qquad
		}{
			\expTypJud{\TtypEnv}{\valK \xname = \e_1; \e_2}{\type_2}
		}
	\end{array} \\
\end{array}$}}
\caption{Typing of \XC\ expressions.} \label{fig:pfc:typing}
\end{figure*}
A \emph{typing environment} $\TtypEnv$ is a set of \emph{assumptions} $\aname : \typescheme$ where the assumption subject $\aname$ can be either a variable, a built-in function or data constructor.
We assume that sensors $\sname$ have types $() \to \type$ where $\FTV{\type} = \emptyset$.
In the typing of programs, an \emph{initial typing environment} $\TtypEnv_0$ declares a (unique) type scheme for every available data constructor and built-in function.
This $\TtypEnv_0$ is then extended with (unique) assumptions for bounded variables encountered while typing sub-expressions of the program. We write $\TtypEnv(\aname)$  for the unique type scheme $\aname$ in $\TtypEnv$.

We specify typing of expressions via judgements $\expTypJud{\TtypEnv}{\e}{\type}\,$
which read ``expression $\e$ has type $\type$ under assumptions $\TtypEnv$''.
Following~\cite{FJ}, multiple overbars are expanded together (e.g., $\expTypJud{\TtypEnv}{\overline\e}{\overline\type}$  stands for $\expTypJud{\TtypEnv}{\e_1}{\type_1}$,  $\cdots$,  $\expTypJud{\TtypEnv}{\e_n}{\type_n}$).
Typing rules for expressions are syntax-directed.
All rules are standard.

\subsection{Devices, Space, Time and Physical Environment} \label{sec:space-and-time}


So far every execution of a given \XC\  program always exhibits the same behavior:  either it produces a value (always the same) or it always non-terminates.
 Now let's spice things up by bringing devices, space, time and physical environment into the picture. Namely, we consider
     a collection of spatially situated \emph{devices},
generally equipped with \emph{sensors} and \emph{actuators}.
These devices  repeatedly compute the same  \XC\ program, which may contain built-in's for accessing sensor values, like those listed in \Cref{fig:standardSensors}.

 \begin{figure*}[tb]
\begin{scriptsize}
\centerline{
\begin{tabular}{@{\hspace{7pt}}c@{\hspace{7pt}}|@{\hspace{7pt}}c@{\hspace{7pt}}|@{\hspace{7pt}}l@{\hspace{7pt}}}
\textsc{Name} & \textsc{Type scheme} & \textsc{Description}
\\\hline
\multicolumn{3}{l}{\textbf{Sensors used in examples:}}
\\\hline
$\uidK$ & $() \to \tnum$ & \emph{Unique device identifier}
\\\hline
$\gpsK$ & $() \to \tpairOf{\tnum}{\tnum}$ & \emph{GPS latitude and longitude}
\\\hline
$\timeK$ & $() \to \tnum$ & \emph{Epoch Unix Timestamp (seconds since Jan 01 1970 (UTC))}
\\\hline
$\temperatureK$ & $() \to \tnum$ & \emph{Temperature (Celsius degrees) sensed by the device}
\\ \hline
\end{tabular}
}
\end{scriptsize}
\caption{\XC: name, type scheme and description of  built-in functions for some sensors.} \label{fig:standardSensors}
\end{figure*}

Device behaviour is modelled through a notion of \emph{(execution) round} or \emph{firing},
 whereby a device independently ``fires'' -- i.e., 
 gathers sensor data and then
 ``atomically executes'' an  \XC\ program.
Executing a terminating \XC\ program
 results in the production of an \emph{output} (the program's  value),
 which may be used to describe actuations,
 before waiting to fire again -- sometimes we say a device ``wakes up'', fires, and then ``goes back to sleep''.
 
All the devices in the collection execute  \emph{the same}
program. However, each device, at each firing, based on the value of sensors, can exhibit a different execution of the  program.
Such firings may be scheduled at comparable periodic intervals on
all
 devices but there is no such assumption in general (every device may have its own scheduling of firings).
Indeed, a device may run out of battery and never wake up again, or it can restart after a long time if the battery gets charged, or it can go into a non terminating execution of the program
and ``disappear'' (a reboot, which aborts the execution and enables another firing, may happen). Therefore, rounds  are entirely asynchronous.

For instance, consider a collection of devices equipped with a screen that displays the value produced by the last completed firing.
If the devices run the program
$\pairK(\pairK(\timeK(),\gpsK()),\temperatureK())$, then each device that has fired at least once will display a value of the form $\PairK(\PairK(\lvalue_1,\PairK(\lvalue_2,\lvalue_3)),$ $\lvalue_4)$ of type $\tpairOf{\tpairOf{\tnum}{\tpairOf{\tnum}{\tnum}}}{\tnum}$, 
where: $\lvalue_1$ is the stating time of the last completed firing;  $\lvalue_2$ and $\lvalue_3$ are the GPS latitude and longitude of the device at the beginning of such firing, respectively; and $\lvalue_4$ is the temperature sensed by the device at the beginning of such firing.

\subsection{Proximity-based Interactions} \label{sec:communication}


So far, if we consider each firing as an event, happening on a spatially situated device at a given time, then the execution of a terminating \XC\ program results in a countable sets of independent events, where each event is labeled by the value produced by the program (if a given firing does not terminate, then it does not produce an event).
Now let's add more spice by bringing proximity-based interaction into the picture. Namely, we consider devices that may communicate asynchronously  with \emph{neighbour} devices by exchanging \emph{messages} and by considering \XC\ programs which may use the $\exchangeK$ built-in for communication
and other built-in's for accessing information received from neighbours, like those listed in \Cref{fig:communicationBuitins}.

 \begin{figure*}[tb]
\begin{scriptsize}
\centerline{
\begin{tabular}{@{\hspace{7pt}}c@{\hspace{7pt}}|@{\hspace{7pt}}c@{\hspace{7pt}}|@{\hspace{7pt}}l@{\hspace{7pt}}}
\textsc{Name} & \textsc{Type scheme} & \textsc{Description}
\\ \hline
\multicolumn{3}{l}{\textbf{Communication:}}
\\\hline
$\exchangeK$ & $(\ltype, (\tfieldOf{\ltype},\tfieldOf{\ltype}) \to \tpairOf{\type}{\tfieldOf{\ltype}}) \to \type$ & \emph{Exchanges messages}
\\ \hline
\multicolumn{3}{l}{\textbf{Neighbouring value manipulation:}}
\\\hline
$\foldK$ & $((\ltype,\ltypealt) \to \ltype, \tfieldOf{\ltypealt}, \ltype) \to \ltype$ & \emph{Folding of a neighbouring value}
\\\hline
$\selfK$ & $(\tfieldOf{\ltype}) \to \ltype$ & \emph{Extract the self-message}
\\ \hline
$\updateSelfK$ & $(\tfieldOf{\ltype},\ltype) \to \tfieldOf{\ltype}$ & \emph{Update the self-message}
\\ \hline
$\updateDefK$ & $(\tfieldOf{\ltype},\ltype) \to \tfieldOf{\ltype}$ & \emph{Update the default-message}
\\\hline
\multicolumn{3}{l}{\textbf{Sensors used in examples:}}
\\\hline
$\nbrDistK$ & $\tfieldOf{\tnum}$ & \emph{Distance estimates to neighbours}
\\ \hline
\end{tabular}
}
\end{scriptsize}
\caption{\XC: name, type scheme and description of  built-in functions for  communication.} \label{fig:communicationBuitins}
\end{figure*}
 
\subsubsection{Messages and Neighbours} 
During a firing a device may schedule messages. Such messages will be sent to neighbours at the end of the firing. Each device collects the received messages in a buffer, keeping only the \emph{most recent} message per neighbour. Moreover,
 messages exceeding a certain configurable lifetime  \emph{expire} (i.e., they are removed from the buffer).
Our formalization assumes that the order of messages from a same sender is preserved but,
other than that, it makes \emph{very few assumptions on messages}. In particular:
	\emph{(i)}
	a device may be too far to send a message to \correction{a (former) neighbour};
	\emph{(ii)}
	a message may get lost;
	\emph{(iii)}
	a device may disappear or fail;
	\emph{(iv)}
	a device may reboot, losing its queue of received messages;
	\emph{(v)}
	a device may deem messages from some devices to be expired.
Notably, \XC\ \correction{abstracts from} those cases:
when a device fires, it finds some messages from
(the most recent available firings of) some other devices, and a device firing does not remove messages from the buffer, allowing
messages to persist across firings.
The devices for which a message is available in a certain firing are \emph{the neighbours}
for that firing. During a firing, a device \correction{may} schedule a message to itself -- such message will be processed in the next firing of the device (if the message does not expire and the device does not reboot before the next firing).

\begin{figure}[t]
\centering
\vspace{-10mm}
  \includegraphics[width=0.45\textwidth]{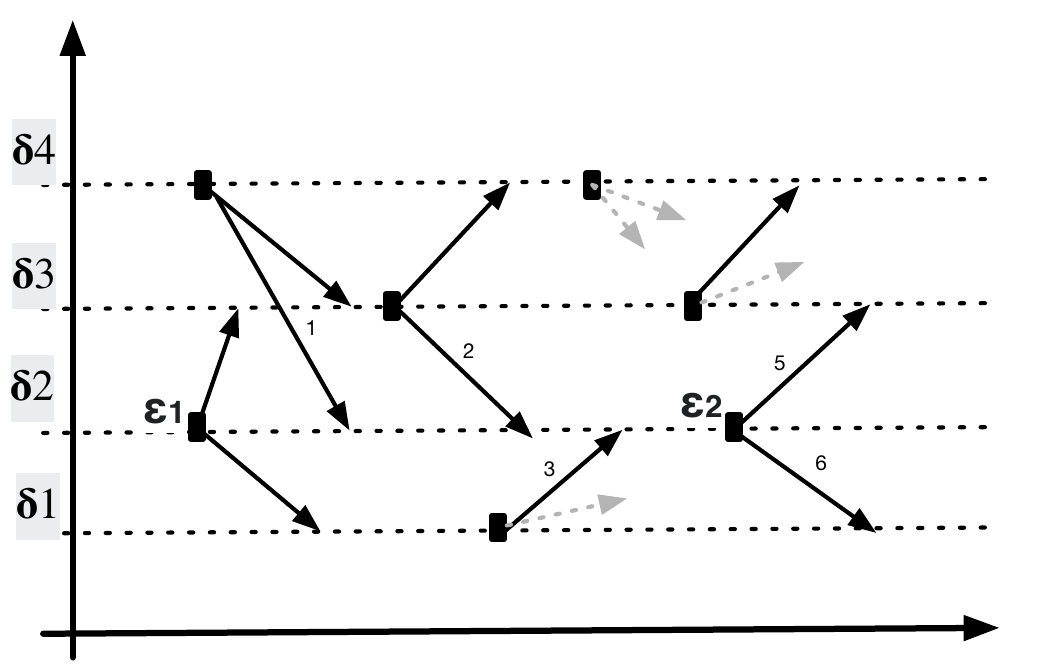}
  \caption{\XC: representation of a system execution (taken from \cite{Audrito-et-al:JSS-2024}).}\label{fig:sys-model}
\end{figure}

\Cref{fig:sys-model} illustrates a sample system execution involving four devices, labeled $\deviceId_1$ through $\deviceId_4$.
During the observed interval, $\deviceId_2$ activates twice, carrying out two computation events, denoted $\eventId_1$ and $\eventId_2$.
Messages shown with gray arrows represent transmissions that fail and are never delivered.
In the second computation $\eventId_2$, device $\deviceId_2$ processes three messages received during its sleep phase, originating from $\deviceId_4$, $\deviceId_3$, and $\deviceId_1$.
Following this computation, $\deviceId_2$ transmits messages to both $\deviceId_3$ and $\deviceId_1$.

\subsubsection{Neighbouring Values}
A neighbouring value (\emph{\excvalue}) is
a map $\underline{\anyvaluealt}$ which associates device identifiers $\deviceId_i$ with literal values $\lvalue_i$ and, moreover, it features a \emph{default} literal value $\lvalue$. It is written $\lvalue[\deviceId_1 \mapsto \lvalue_1, \ldots, \deviceId_n \mapsto \lvalue_n]$ $(n\ge 0)$, to be read ``the {\excvalue} $\underline{\anyvaluealt}$ is $\lvalue$
everywhere (i.e., for all neighbours) except for devices $\deviceId_1, \ldots, \deviceId_n$
with values $\lvalue_1, \ldots, \lvalue_n$''.
\Excvalues\ cannot occur in the source code of programs: they arise at runtime. Each \excvalue\ is used to describe the (set of) literal values received from and sent to neighbours.
The devices with an associated entry in the \excvalues{} manipulated by a device $\deviceId$ are usually
a subset of all devices,
e.g., because some devices are too far from $\deviceId$ to provide a value
or the last provided value has expired.
The default is used when the literal  value for a device is not available (e.g., if the device manipulating the \excvalue\ just appeared/rebooted and, therefore, there is no previous firing that scheduled a message to itself).

To illustrate the use of \excvalues, while referring to \Cref{fig:sys-model}: when $\deviceId_2$ wakes up for computation $\eventId_2$, it may construct an \excvalue\ $\underline{\anyvaluealt} = 0[\correction{\deviceId_4 \mapsto 1},\deviceId_3 \mapsto 2, \correction{\deviceId_1 \mapsto 3}]$, which aggregates scalar values received during its sleep phase from $\deviceId_4$, $\deviceId_3$, and $\deviceId_1$, namely \correction{1, 2, and 3}, respectively. Any device not explicitly listed contributes the default value $0$.
Following this computation, $\deviceId_2$ might issue outgoing messages captured by the \excvalue\ $\underline{\anyvaluealt}' = 0[\deviceId_3 \mapsto 5,\correction{\deviceId_1 \mapsto 6}]$,
meaning $\deviceId_3$ receives 5, $\deviceId_1$ receives \correction{6}, and all other recipients—such as devices that have just joined—receive the default value 0.

The notation $\underline{\anyvaluealt}(\deviceId')$ denotes the explicit value $\lvalue'$ assigned to $\deviceId'$ if present in $\underline{\anyvaluealt}$, or the default value $\lvalue$ otherwise, highlighting the interpretation of \excvalues\ as maps with a fallback. For example, we have $\underline{\anyvaluealt}'(\deviceId_1) = 6$ and $\underline{\anyvaluealt}'(\deviceId_2) = 0$.
For clarity, in code snippets we underline variables that carry neighbour-specific values. Likewise, we underline a primitive type $\tfieldOf{\ltype}$ to indicate the type of an \excvalue\ expression like $\underline{\anyvaluealt} = \lvalue[\deviceId_1 \mapsto \lvalue_1, \ldots, \deviceId_n \mapsto \lvalue_n]$, where each $\lvalue$ and $\lvalue_i$ is of type $\ltype$.

\paragraph{Nvalues generalise literal values}
Any literal value $\lvalue$ can be implicitly promoted to an \excvalue\ of the form $\lvalue[]$, where all devices are assigned the same default value. This allows the formal treatment of literal values and \excvalues\ to be unified.
Moreover, operations defined over literal values are automatically lifted to operate over \excvalues\ in a pointwise fashion, meaning the function is applied to each mapped value individually.
For example, given
$\underline{\anyvaluealt}_1 =   0[\deviceId_1\!\mapsto\!1,\deviceId_2\!\mapsto\!2]$ and
$\underline{\anyvaluealt}_2 =   2[\deviceId_2\!\mapsto\!1]$,
we have
$\underline{\anyvaluealt}_3 = \underline{\anyvaluealt}_1 + \underline{\anyvaluealt}_2 =   2[\deviceId_1\!\mapsto\!3,\deviceId_2\!\mapsto\!3]$. Note that $\deviceId_1\!\mapsto\!3$ in $\underline{\anyvaluealt}_3$
is due to the fact that $\correction{\deviceId_1\!\mapsto\!1}$ in $\underline{\anyvaluealt}_1$ and
$\deviceId_1$ has default value 2 in $\underline{\anyvaluealt}_2$.
Moreover, thanks to the automatic promotion of literals to \excvalues, expressions like $\underline{\anyvaluealt}_1 + 1$ are also valid and interpreted as
$\underline{\anyvaluealt}_1 + 1 =   0[\deviceId_1\!\mapsto\!1,\deviceId_2\!\mapsto\!2] + 1
= 1[\deviceId_1\!\mapsto\!2,\deviceId_2\!\mapsto\!3]$, where the scalar value $1$ is lifted to the \excvalue\ $1[]$ and then added pointwise.

\paragraph{Operations on nvalues}

In addition to supporting pointwise operations, \excvalues\ can also be reduced via folding, akin to list folding. This is done using the built-in construct $\foldK(f: (A,B) \to A, \underline{\anyvaluealt}: \tfieldOf{B}, \lvalue: A): A$, where the function $f$ is applied iteratively to the values associated with neighbouring devices in the field $\underline{\anyvaluealt}$ — excluding the value for the device executing the computation — starting from the initial value $\lvalue$.

Consider, for example, device $\deviceId_2$ executing a fold over its neighbours $\bp{\deviceId_1, \deviceId_3}$. If we define $\underline{\anyvaluealt}_1 = 0[\deviceId_1 \mapsto 1,\deviceId_2 \mapsto 2]$ as before, then the expression $\foldK(+, \underline{\anyvaluealt}_1, 10)$ evaluates to $10 + \underline{\anyvaluealt}_1(\deviceId_1) + \underline{\anyvaluealt}_1(\deviceId_3) = 10 + 1 + 0$.
Since the semantics of \excvalues\ do not rely on any specific ordering of device identifiers, it is standard to assume that the folding function $f$ is both associative and commutative.

\begin{example}[Distance estimation] \label{exa:dist:body}
A device can estimate its distance to a target node in the network by relying on distance estimates $\underline{\mathtt{n}}$ communicated by its neighbours.
This is achieved by computing the minimum of the neighbours' values $\underline{\mathtt{n}}$, each incremented by a local estimate of the distance to that neighbour, captured in $\underline{\mathtt{senseDist}}$ (typically provided by a sensor on the device).
The combination is performed using \lstinline|nfold|, which starts from the value \lstinline|Infinity| and applies the \lstinline|min| function over all adjusted neighbour estimates:
\begin{lstlisting}[mathescape]
def distanceEstimate($\underline{\mathtt{n}}$) { // type: $(\tfieldOf{\tnum}) \to \tnum$
  nfold(min, $\underline{\mathtt{n}}$ + $\underline{\mathtt{senseDist}}$, Infinity)
}
\end{lstlisting}
Here, the fields $\underline{\mathtt{n}}$ and $\underline{\mathtt{senseDist}}$ are summed pointwise across neighbours. That is, for each neighbour $\deviceId$, the device computes the sum $\underline{\mathtt{n}}(\deviceId) + \underline{\mathtt{senseDist}}(\deviceId)$ as an individual estimate. The final result is the smallest of these summed values, or \lstinline|Infinity| if no values are available.
\end{example}

Further built-in operations for \excvalues\ are considered available. Among them $\selfK(\underline{\anyvaluealt} : \tfieldOf{\ltype}) : \ltype$, which extracts the value associated with the current device $\deviceId$ from the field $\underline{\anyvaluealt}$—that is, it returns $\underline{\anyvaluealt}(\deviceId)$.
Furthermore, $\updateSelfK(\underline{\anyvaluealt} : \tfieldOf{\ltype}, \lvalue : \ltype) : \tfieldOf{\ltype}$, which produces a new \excvalue\ by updating $\underline{\anyvaluealt}$ at the key corresponding to the self device, replacing the previous value (if any) with $\lvalue$, while leaving all other entries unchanged.
This operation corresponds to a substitution in a defaulted map, and is denoted using the syntax $\underline{\anyvaluealt}[\deviceId \mapsto \lvalue]$. More generally, expressions like $\lvalue[\deviceId_1 \mapsto \lvalue_1, \ldots]$ represent maps that assign $\lvalue_i$ to $\deviceId_i$ and use $\lvalue$ as the fallback value for all other devices, effectively describing updates over a uniform default map.

\subsubsection{The $\exchangeK$ built-in function} \label{ssec:communication}

\XC\ features a single communication primitive
$\dExchange{\e_i}{\underline{\oname},\underline{\nname}}{\e_r}{\e_s}$ which de-sugars to
$\pExchange{\e_i}{\underline{\oname},\underline{\nname}}{\e_r}{\e_s}$
\correction{which models a collective computation through a local starting value $\e_i$ and an \emph{update function}. This function computes an nvalue $\e_r$ to return and an nvalue $\e_s$ to send to neighbours, given an \emph{old} value $\oname$ (the nvalue of outgoing messages computed in the previous round of the current device) and a \emph{neighbour} value $\nname$ (the nvalue of incoming messages for the current device, that neighbouring devices computed in their most recent rounds).
Such expression}
is evaluated as follows:
\emph{(i)} the device evaluates expression $\e_i$ to a a literal value $\lvalue_i$ (the \emph{initial} value);
\emph{(ii)} it substitutes variable $\underline{\oname}$ with the \excvalue\ $\underline{\anyvaluealt}_o$ computed in the previous round by the local device for $\e_s$ (using $\lvalue_i$ instead, if it is the first round computing this exchange construct for the current device);
\emph{(iii)} it substitutes variable $\underline{\nname}$ with the \excvalue\ $\underline{\anyvaluealt}_n$ of messages
received from the neighbours for this exchange, using $\lvalue_i$ as default;
\emph{(iv)}
the whole exchange expression returns the (neighbouring or literal) value $\anyvalue_r$ obtained from the evaluation of $\e_r$;
\emph{(v)} evaluate $\e_s$ to a \excvalue\ $\underline{\anyvaluealt}_s$
consisting of literal values to be sent to neighbour devices $\deviceId'$,
that will use their corresponding $\underline{\anyvaluealt}_s(\deviceId')$ to build their $\underline{\anyvaluealt}_n$ values for their following execution round.

Since most times expressions $\e_r$ and $\e_s$ coincide, we provide
$\sExchange{\e_i}{\underline{\oname},\underline{\nname}}{\e}$ as a shorthand for $\vExchange{\e_i}{\underline{\oname},\underline{\nname}}{\e}$.
A further useful communication pattern is to retrieve neighbours' values for an expression, which can be accomplished through the defined function
$\nbrK(\e_i, \e_s) = \dExchange{\e_i}{\underline{\oname},\underline{\nname}}{\underline{\nname}}{\e_s}$. In $\nbrK(\e_i, \e_s)$, the value of expression $\e_s$ is sent to neighbours, and the values received from them (gathered in $\underline{\nname}$ together with the default from $\e_i$) are returned as a \excvalue,
thus providing a view on neighbours' values of $\e_s$.
Another commonly used pattern is state preservation across rounds, which can be obtained through the defined function $\oldK(\e_i, \e_s) = \dExchange{\e_i}{\underline{\oname},\underline{\nname}}{\underline{\oname}}{\e_s}$. In this case, the value of expression $\e_s$ is shared with neighbours but only relevant for the current device, which will use it in its following round as the overall result of the exchange expression.

A key aspect of \XC's\ expressiveness lies in the ability of \emph{exchange} (and thus \emph{nbr} and \emph{old}) to assign distinct values to different neighbours. This capability enables tailored interactions between devices, which is essential for many coordination patterns, as illustrated in the following example.

\begin{example}[Ping-pong counter]\label{exa:conn}
The function below computes a neighbouring field of ``connection counters'', mapping each neighbour to the number of times a bidirectional connection has been formed with it.
\begin{lstlisting}[mathescape]
def ping-pong() { // type: $() \to \tfieldOf{\tnum}$
  exchange( 0, ($\underline{\mathtt{o}},\underline{\mathtt{n}}$) => retsend $\underline{\mathtt{n}}$ + 1 )
}
\end{lstlisting}
Upon each evaluation of \lstinline|ping-pong|, a device retrieves a field $\underline{\anyvaluealt}$ that assigns to each neighbour its current connection count—defaulting to $0$ for newly connected peers.
Then, using $\underline{\mathtt{n}} + 1$ (where $\underline{\mathtt{n}}$ stands for $\underline{\anyvaluealt}$), the device increments every neighbour's counter, including those just connected (who thus get mapped to $1$).
This updated field, $\underline{\anyvaluealt} + 1$, is both shared with neighbours and returned as the function's result.

While the connection persists, each device continuously receives its neighbour’s counter, increments it, and sends it back—effectively counting the ping-pong message exchanges.
If the connection drops (and related messages expire), the counter resets to $0$. Then, it may restart incrementing if the connection is re-established later.
Crucially, the function maintains a separate counter for each neighbour, enabling individualized network tracking through the ability to send neighbour-specific values.
\end{example}

The previous example provided means to track the persistence of bidirectional connections. In the following example, we present different connection counter, that is tailored to measure the persistence of a unidirectional connection.

\begin{example}[Unidirectional Connection Counter]\label{exa:uniconn}
The \emph{uniconn-count} function defined below generates an \excvalue\ representing inbound connection counters, assigning to each neighbour the number of times a message has been successfully received from them.
\begin{lstlisting}[mathescape]
def uniconn-count() {
  exchange( 0, ($\underline{\mathtt{o}},\underline{\mathtt{n}}$) => retsend $\underline{\mathtt{o}}$ + updateDef(1,0) ) )
}
\end{lstlisting}
Note that \emph{uniconn-count} uses only the argument $\underline{\mathtt{o}}$ and ignores $\underline{\mathtt{n}}$.
The exchange begins with an initial value of $0$. In each round, the update step increments the value associated with all current neighbours in the field $\underline{\mathtt{o}}$, while keeping the default value unchanged.
This behavior is implemented using the \emph{updateDef} function, which—when applied to inputs $1$ and $0$—yields a neighbouring value with a default of $0$ and assigning $1$ to all present neighbours.
As a result, this mechanism counts how many consecutive rounds a given device has remained a neighbour. The counter resets to $0$ whenever a previously connected neighbour disappears from the current view.

Since neighbouring devices are those that have successfully transmitted a message to the current device in recent rounds, this function serves as a proxy for assessing the quality of inbound connectivity between the device and its peers.
Importantly, unlike the \emph{ping-pong} counter, it does not reflect the quality of outgoing communication: even if the current device fails to send messages, the counters still progress as long as it continues to receive.
\end{example}

As a final example, we now present the \emph{self-organising distance} algorithm, which highlights how \emph{exchange} works in combination with \emph{nfold} to achieve distributed coordination.

\begin{example}[Self-organising distance] \label{exmp:distance}
Calculating the shortest distance from each device to a designated set of \emph{source} devices produces what is known as a \emph{gradient} \cite{audrito2017ULT}. Gradients represent a foundational self-organisation pattern, useful for tasks such as approximating long-range distances and guiding data propagation along efficient paths.
The \lstinline|distanceTo| function implements this mechanism through a distributed variant of the Bellman-Ford algorithm\cite{DasguptaCDC16}.
\begin{lstlisting}[mathescape]
def distanceTo(src) { // type: $(\tbool) \to \tnum$
  exchange( Infinity, ($\underline{\mathtt{n}}$) => retsend mux(src, 0, distanceEstimate($\underline{\mathtt{n}}$)) )
}
\end{lstlisting}
When executed repeatedly across a network, the function converges to stable distance values from source nodes—i.e., those where \lstinline|src| evaluates to true.
The core \lstinline|exchange| expression refines the local estimate by:
\emph{(i)} defaulting to \lstinline|Infinity| when no input is available,
\emph{(ii)} assigning a distance of zero on source devices, and
\emph{(iii)} for all other nodes, computing the minimum among neighbours’ distances plus the sensed inter-device distances (see Example~\ref{exa:dist:body}).
If the resulting estimate is $d$, this value is returned and also shared with neighbours using a uniform field mapping each peer to $d$.

The operator \lstinline|mux| is used to ensure both branches of the conditional are evaluated: even though source devices return $0$, they must still compute and share their neighbour field $\underline{\mathtt{n}}$ via \lstinline|distanceEstimate|.
This design supports dynamic adaptation: any change in the network topology—due to failures, movement, or devices joining/leaving—directly alters the domain of $\underline{\mathtt{n}}$, triggering recalculation at the local and global levels.
\end{example}

\tikzset{-,
  ev/.style={circle,draw, inner sep=1.8pt, minimum size=2pt, outer sep=0.5mm},
  point/.style={circle,fill=black, inner sep=1.8pt, minimum size=2pt, outer sep=0.5mm},
  pointp/.style={rectangle,fill=black, inner sep=2pt, minimum size=2pt, outer sep=0.5mm},
  event/.style={circle,fill=red, inner sep=3pt, minimum size=2pt, outer sep=0.5mm},
  past/.style={event, fill=ddarkgreen},
  present/.style={event,fill=gray},
  future/.style={event,fill=blue},
  locality/.style={circle,fill=white, inner sep=2pt, minimum size=2pt, outer sep=0.5mm},
  locality1/.style={locality,fill=orange!50},
  locality2/.style={locality,fill=gray!50},
  locality3/.style={locality,fill=blue!40},
  locality4/.style={locality,fill=black!30!green},
  locality5/.style={locality,fill=purple!50},
  d/.style={circle,fill=gray, inner sep=2pt, minimum size=2pt},
  da/.style={circle,draw,label={[left]$1$}},
  db/.style={circle,draw,label={[left]$2$}},
  dc/.style={circle,draw,label={[left]$3$}},
  lbl/.style={font=\footnotesize},
  state/.style={circle,draw},
  statein/.style={circle,draw,line width={2pt}},
  stateout/.style={circle,draw,line width={1pt},fill=black!16},
  leadstov/.style={draw,decorate,decoration={snake, amplitude=1.95mm, post=lineto, post length=2mm, segment length=1cm, pre length=0.4cm},->,>=stealth'},
  leadsto/.style={draw,decorate,decoration={snake, amplitude=0.35mm, post=lineto, post length=2mm, segment length=1.3mm},->,>=stealth'}, 
  leadstopast/.style={leadsto,draw=ddarkgreen},
  leadstopresent/.style={leadsto,draw=black},
  leadstofuture/.style={leadsto,draw=blue},
  frastagliato/.style={draw,decorate,decoration={snake, amplitude=0.45mm, post=lineto, post length=0mm, segment length=2.3mm},->,>=stealth'},
  appgraph/.style={rectangle,draw,minimum size=1.5cm},
  appgraphc/.style={ellipse,draw,minimum size=1.5cm},
  program/.style={rectangle,draw,radius=0.1cm},
  scheduler/.style={rounded rectangle,draw,radius=0.1cm,font=\small},
  evpast/.style={green},
  evpresent/.style={red},
  evfuture/.style={blue},
  evconc/.style={gray},
}

\subsection{Compositionality through alignment} \label{ssec:alignment}

When a program contains multiple \lstinline|exchange| expressions, \XC\ guarantees correct message delivery with a mechanism called \emph{alignment}. This ensures that values exchanged in one round are received by the corresponding exchange expression in the next round—specifically, the one in the same AST position and stack frame. This alignment holds even in the presence of branching, function calls, or recursion.

\begin{example}[Neighbour average]
\label{ex:average}
The \lstinline|average| function computes a weighted average of a value over a device’s immediate neighbours.
\begin{lstlisting}[mathescape]
def average(weight, value) { // type: $(\tnum) \to \tnum$
  val totW = nfold(+, nbr(0, weight), weight); %\label{totW}%
  val totVl = nfold(+, nbr(0, weight*value), weight*value);%\label{totV}%
  totV / totW
}
\end{lstlisting}
Line~\ref{totW} calculates the total weight by exchanging neighbour weights via \lstinline|nbr(0, weight)| and summing them with \lstinline|nfold|, including the local device by using \lstinline|weight| as the base. Line~\ref{totV} performs a similar operation for the weighted values, summing \lstinline|weight * value| across neighbours. The result is returned as \lstinline|totV / totW|.
The function involves two \lstinline|nbr| calls (invoking \lstinline|exchange|) with messages of type $\tnum$. \XC\ guarantees correct message alignment, so that each exchange receives only the corresponding data (weights or values) from neighbours, without interference between them.
\end{example}

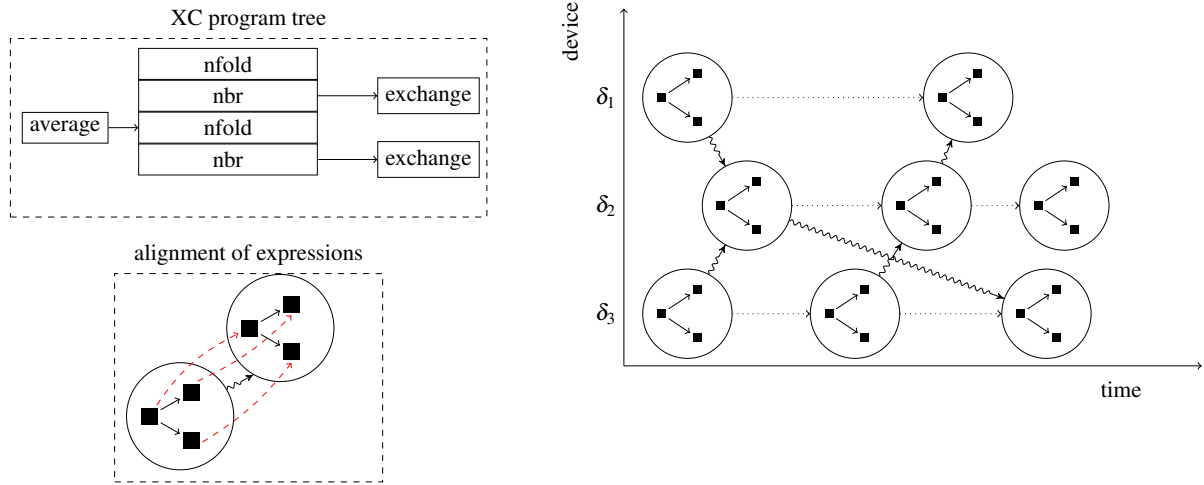
\begin{figure*}
\centering
\resizebox{1.0\textwidth}{!}{
\begin{tikzpicture}[node distance=2cm]
\newcommand{\drawgraphin}[2]{\drawappgraphin{#1}{0.4cm}{0.4cm}{0.25}{pointp}{#2}}
\makeatletter
\newcommand{\drawappgraphin}[6]{
\node[#5] (#1p1) [right=0.2cm of #1.west] {};

		\node[#5] (#1p3) [right=2*#2 of #1.west,yshift=#3] {};
		\node[#5] (#1p4) [right=2*#2 of #1.west,yshift=-#3] {};
		\draw[->] (#1p1) -- (#1p3);
		\draw[->] (#1p1) -- (#1p4);
}
\makeatother

\node[rectangle,draw,dashed,black,minimum width=8cm,minimum height=3cm] (legend) at (-6cm,0cm) [label={[align=center]above:{\XC{} program tree
}}] {};

\node[draw] (avg) [right=0.2cm of legend.west] {average};
\node[draw,minimum width=3cm] (avgbody1) [below right=0.0cm and 0.5cm of avg] {nbr};
\node[draw,minimum width=3cm] (avgbody2) [above=0cm of avgbody1] {nfold};
\node[draw,minimum width=3cm] (avgbody3) [above=0cm of avgbody2] {nbr};
\node[draw,minimum width=3cm] (avgbody4) [above=0cm of avgbody3] {nfold};

\node[draw] (exch1) [right=1cm of avgbody1] {exchange};
\node[draw] (exch2) [right=1cm of avgbody3] {exchange};
\draw[->] (avg) -- (avgbody2.180);
\draw[->] (avgbody1.east) -- (exch1.west);
\draw[->] (avgbody3.east) -- (exch2.west);

\node[] (d1l) [above=0.2cm] {$\delta_1$};
\node[] (d2l) [below=1.2cm of d1l] {$\delta_2$};
\node[] (d3l) [below=1.2cm of d2l] {$\delta_3$};

\node[appgraphc] (d1e1) [right=0.3cm of d1l] {};
\node[appgraphc] (d1e2) [right=3.2cm of d1e1] {};
\node[appgraphc] (d2e1) [right=1.3cm of d2l] {};
\node[appgraphc] (d2e2) [right=1.5cm of d2e1] {};
\node[appgraphc] (d2e3) [right=0.8cm of d2e2] {};
\node[appgraphc] (d3e1) [right=0.3cm of d3l] {};
\node[appgraphc] (d3e2) [right=1.3cm of d3e1] {};
\node[appgraphc] (d3e3) [right=1.7cm of d3e2,
] {};

\draw[->] (0.3,-4) -- node[xshift=-0.9cm,yshift=2.7cm,rotate=90]{devices} (0.3,2);
\draw[->](0.3,-4) -- node[yshift=-0.4cm,xshift=3.5cm]{time} (10,-4);

\drawgraphin{d1e1}{0}
\drawgraphin{d1e2}{1}
\drawgraphin{d2e1}{1}
\drawgraphin{d2e2}{0}
\drawgraphin{d2e3}{0}
\drawgraphin{d3e1}{0}
\drawgraphin{d3e2}{0}
\drawgraphin{d3e3}{0}

\draw[->,dotted] (d1e1) -- (d1e2);
\draw[->,dotted] (d2e1) -- (d2e2);
\draw[->,dotted] (d3e1) -- (d3e2);
\draw[->,dotted] (d3e2) -- (d3e3);
\draw[leadsto] (d1e1) -- (d2e1);
\draw[leadsto] (d3e1) -- (d2e1);
\draw[leadsto] (d2e1) -- (d3e3);
\draw[leadsto] (d2e2) -- (d1e2);
\draw[->,dotted] (d2e2) -- (d2e3);
\draw[leadsto] (d3e2) -- (d2e2);

\node[rectangle,draw,dashed,black,minimum width=4.5cm,minimum height=3.5cm] (legend) at (-6cm,-4.2cm) [label={[align=center]above:{alignment of 
expressions}}] {};
\node[appgraphc,minimum size=1.8cm] (n1) [right=0.2cm of legend.196, yshift=0.0cm] {};
\node[appgraphc,minimum size=1.8cm] (n2) [above right=0.2cm and 0.4cm of n1] {};
\drawappgraphin{n1}{0.45cm}{0.4cm}{0.25}{pointp,inner sep=4pt}{1};
\drawappgraphin{n2}{0.45cm}{0.4cm}{0.25}{pointp,inner sep=4pt}{1};
\draw[leadsto] (n1.30) to (n2.240);
\draw[->,red,dashed] (n1p1) to [bend left=15] (n2p1);
\draw[->,red,dashed] (n1p4.350) to [bend right=10] (n2p4.270);
\draw[->,red,dashed] (n1p3.60) to [bend right=10] (n2p3.270);

\end{tikzpicture}
}
\caption{\XC{} alignment mechanism for \Cref{ex:average}.
}
\label{fig:alignment-pictorial}
\end{figure*}

\Cref{fig:alignment-pictorial} illustrates this process.
The top-left depicts the stack frames and AST of the \XC{} program from Example~\ref{ex:average}, focusing on \lstinline|exchange|, \lstinline|nfold|, and related functions.
The top-right shows a system execution: dotted arrows link successive rounds of the same device, while curly arrows represent inter-device messages. Each round re-evaluates the same program structure.
The bottom-left zooms into two rounds on different devices evaluating \lstinline|average|, where matching AST nodes align correctly. Red dashed arrows highlight how corresponding \lstinline|exchange| calls communicate across rounds.

Alignment is essential in \XC\ as it enables \emph{compositional distributed programming}, ensuring that data flows between devices follow the intended logical structure of the code.

\begin{example}[Fire detection] \label{exmp:fire-detection}
The \lstinline|closestFire| function estimates the distance to the nearest potential fire by combining \lstinline|average| and \lstinline|distanceTo|, using sensor inputs \lstinline|temperature| and \lstinline|smoke|.
\begin{lstlisting}[mathescape]
def closestFire(temperature, smoke) {//type:$({\tnum,\tnum}) \to \tnum$
  val trust = nfold(+, 1, 1); %\label{trust}%
  val hot = average(trust, temperature) > 60; %\label{hot}%
  val cloudy = average(trust, smoke) > 10; %\label{cloudy}%
  distanceTo(hot and cloudy) %\label{distFire}%
}
\end{lstlisting}
Line~\ref{trust} computes a trust score proportional to the device’s number of neighbours as \mbox{\lstinline|nfold(+, 1, 1)|.} Lines~\ref{hot} and~\ref{cloudy} use this trust to weight averages of temperature and smoke, checking if they exceed 60°C and 10\%, respectively. Line~\ref{distFire} then computes the distance to nodes where both thresholds are met.
Multiple \lstinline|exchange| calls occur within \lstinline|average| and \lstinline|distanceTo|: thanks to alignment, each receives the correctly matched messages from the same expressions in neighbours’ previous rounds.
\correction{In fact, without alignment, the messages resulting from the two calls to the \lstinline|average| function would be indistinguishable, so that we may use smoke levels to compute the average temperature or vice-versa.}
\end{example}

\XC\ supports conditionals like \lstinline|if (cond) {e1} else {e2}|, with semantics that affect communication. Since exchange alignment depends on AST position and stack frame, only devices taking the \emph{same} branch align and exchange data. Devices in different branches are considered non-aligned and do not share messages or contribute to operations like \lstinline|exchange| or \lstinline|nfold|. This effectively partitions the network: each branch forms an isolated sub-network. 

\begin{example}[Domain-isolated computations]\label{exmp:exch-if}
Consider a network of service requesters and providers, where devices are dynamically divided into two groups: those doing local computation (\lstinline|local|) and those offloading it via \lstinline|gateway| nodes. We want to compute distances to gateways while ignoring local devices. \\

\begin{lstlisting}[mathescape]
// type: $(\tbool, \tbool) \to \tnum$
def distanceToGateways(local, gateway) {
  if (local) { Infinity } else { distanceTo(gateway) }
}
\end{lstlisting}
When \lstinline|local| is true, a device returns \lstinline|Infinity|, effectively acting as an obstacle. Otherwise, it evaluates \lstinline|distanceTo(gateway)|, which includes an exchange. Devices in the \lstinline|local| group skip this computation and don't participate in the exchange.
Now suppose we want \lstinline|local| devices to compute distances to nearby \lstinline|requester|s, while non-\lstinline|local| devices compute distances to \lstinline|gateway|s—each excluding the other group. 
\begin{lstlisting}[mathescape]
// type: $(\tbool, \tbool, \tbool) \to \tbool$
def distanceInServiceProvisioning(local,requester,gateway){
  if (local) { distanceTo(requester) }
  else { distanceTo(gateway) }
}
\end{lstlisting}
Although all calls to \lstinline|distanceTo| occur at the same AST position, they lie in different branches and thus have distinct stack frames. As a result, \XC\ ensures that message exchanges remain separated, preventing unintended cross-group communication.
\end{example}

\begin{figure*}
\centering
\resizebox{1.0\textwidth}{!}{
\begin{tikzpicture}[node distance=2cm]
\newcommand{\drawgraphin}[2]{\drawappgraphin{#1}{0.25cm}{0.4cm}{0.25}{pointp}{#2}}
\makeatletter
\newcommand{\drawappgraphin}[6]{
\node[#5] (#1p1) [right=0.1cm of #1.west] {};

\ifthenelse{\equal{#6}{0}}{ 
		\node[#5] (#1p3) [right=2.0*#2 of #1.west,yshift=#3] {};
		\node[#5] (#1p5) [right=1.2*#2 of #1p3] {};
		\node[#5,fill=white,draw,dotted] (#1p4) [right=2.0*#2 of #1.west,yshift=-#3] {};
		\node[#5,fill=white,draw,dotted] (#1p6) [right=1.2*#2 of #1p4] {};

		\draw[->] (#1p1) -- (#1p3);
		\draw[->,dotted] (#1p1) -- (#1p4);
		\draw[->] (#1p3) -- (#1p5);
		\draw[->,dotted] (#1p4) -- (#1p6);
}{
		\node[#5,fill=white,draw,dotted] (#1p3) [right=2.0*#2 of #1.west,yshift=#3] {};
		\node[#5,fill=white,draw,dotted] (#1p5) [right=1.2*#2 of #1p3] {};
		\node[#5] (#1p4) [right=2.0*#2 of #1.west,yshift=-#3] {};
		\node[#5] (#1p6) [right=1.2*#2 of #1p4] {};

		\draw[->,dotted] (#1p1) -- (#1p3);
		\draw[->] (#1p1) -- (#1p4);
		\draw[->,dotted] (#1p3) -- (#1p5);
		\draw[->] (#1p4) -- (#1p6);
}
}
\makeatother

\node[rectangle,draw,dashed,black,minimum width=16cm,minimum height=3cm] (legend) at (-1cm,0cm) [label={[align=center]above:{\XC{} program tree
}}] {};
\node[draw] (avg) [right=0.5cm of legend.west,yshift=-0.5cm] {distanceInServiceProvisioning};

\node[draw,minimum width=3cm] (dist1) [right=0.0cm and 0.5cm of avg] {distanceTo};
\node[draw,minimum width=3cm] (dist2) [above=0.0cm of dist1] {distanceTo};

\node[draw,minimum width=3cm] (dimpl1) [below right=0cm and 1cm of dist1] {distanceEstimate};
\node[draw,minimum width=3cm] (dimpl12) [above=0cm of dimpl1] {exchange};
\node[draw] (dimpl121) [right=1cm of dimpl1] {nfold};
\node[draw,minimum width=3cm] (dimpl2) [above right=0cm and 1cm of dist2] {distanceEstimate};
\node[draw,minimum width=3cm] (dimpl22) [above=0cm of dimpl2] {exchange};
\node[draw] (dimpl221) [right=1cm of dimpl2] {nfold};

\draw[->] (avg.east) -- (dist1.176);

\draw[->] (dist1.east) -- (dimpl1.west);
\draw[->] (dist2.east) -- (dimpl2.west);
\draw[->] (dimpl2.east) -- (dimpl221.west);
\draw[->] (dimpl1.east) -- (dimpl121.west);

\node[] (d1l) at (-1,-3) {$\delta_1$};
\node[] (d2l) [below=1.2cm of d1l] {$\delta_2$};
\node[] (d3l) [below=1.2cm of d2l] {$\delta_3$};

\node[appgraphc] (d1e1) [right=0.3cm of d1l] {};
\node[appgraphc] (d1e2) [right=3.2cm of d1e1] {};
\node[appgraphc] (d2e1) [right=1.3cm of d2l] {};
\node[appgraphc] (d2e2) [right=1.5cm of d2e1] {};
\node[appgraphc] (d2e3) [right=0.8cm of d2e2] {};
\node[appgraphc] (d3e1) [right=0.3cm of d3l] {};
\node[appgraphc] (d3e2) [right=1.3cm of d3e1] {};
\node[appgraphc] (d3e3) [right=1.7cm of d3e2,
] {};

\draw[->] (-1.6,-7.5) -- node[xshift=-0.5cm,yshift=2.2cm,,rotate=90]{devices} (-1.6,-2);
\draw[->](-1.6,-7.5) -- node[yshift=-0.4cm,xshift=3.5cm]{time} (8,-7.5);

\drawgraphin{d1e1}{0}
\drawgraphin{d1e2}{1}
\drawgraphin{d2e1}{1}
\drawgraphin{d2e2}{0}
\drawgraphin{d2e3}{0}
\drawgraphin{d3e1}{0}
\drawgraphin{d3e2}{0}
\drawgraphin{d3e3}{0}

\draw[->,dotted] (d1e1) -- (d1e2);
\draw[->,dotted] (d2e1) -- (d2e2);
\draw[->,dotted] (d3e1) -- (d3e2);
\draw[->,dotted] (d3e2) -- (d3e3);
\draw[leadsto] (d1e1) -- (d2e1);
\draw[leadsto] (d3e1) -- (d2e1);
\draw[leadsto] (d2e1) -- (d3e3);
\draw[leadsto] (d2e2) -- (d1e2);
\draw[->,dotted] (d2e2) -- (d2e3);
\draw[leadsto] (d3e2) -- (d2e2);

\node[rectangle,draw,dashed,black,minimum width=4.5cm,minimum height=3.5cm] (legend) at (-6cm,-4.2cm) [label={[align=center]above:{partial alignment of 
expressions}}] {};
\node[appgraphc,minimum size=2.0cm] (n3) [right=0.1cm of legend.196, yshift=0.0cm] {};
\node[appgraphc,minimum size=2.0cm] (n4)  [above right=0.05cm and 0.65cm of n1]  {};
\drawappgraphin{n3}{0.3cm}{0.4cm}{0.25}{pointp,inner sep=4pt}{0};
\drawappgraphin{n4}{0.3cm}{0.4cm}{0.25}{pointp,inner sep=4pt}{1};
\draw[leadsto] (n3.30) to (n4.240);
\draw[->,red,dashed] (n3p1.120) to [bend left=40] (n4p1);

\end{tikzpicture}
}
\caption{\XC{} alignment mechanism with conditionals for \Cref{exmp:exch-if}.}
\label{fig:partial-alignment-pictorial}
\end{figure*}
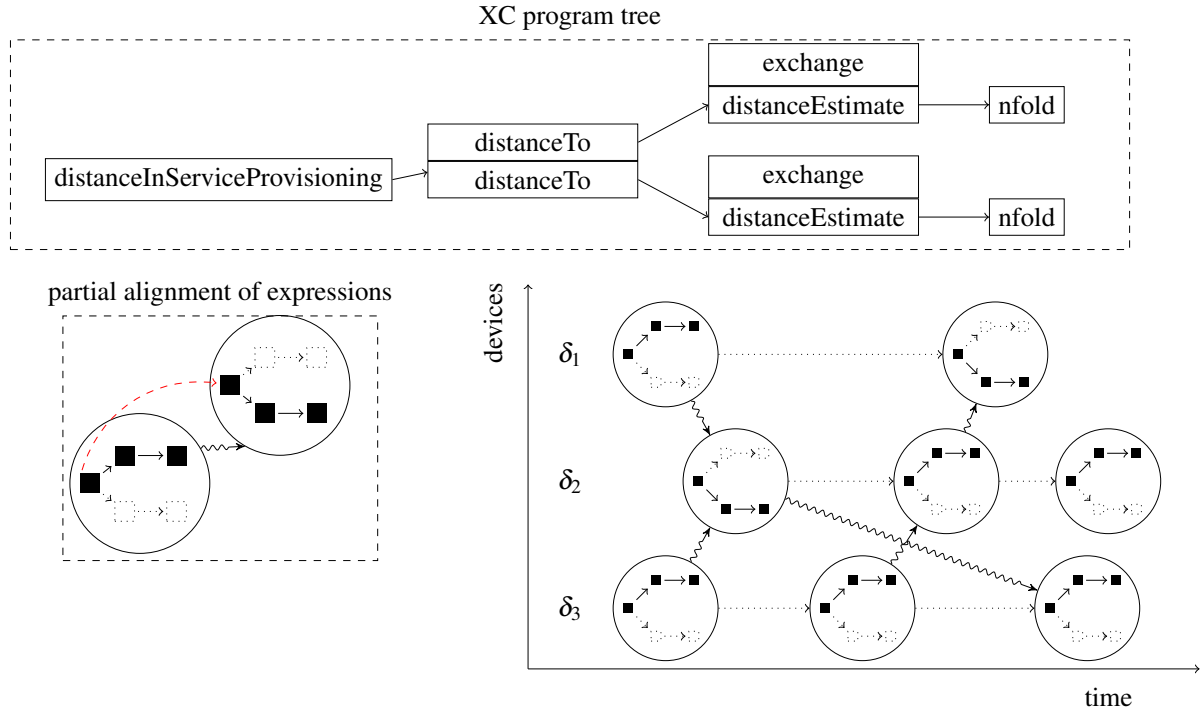

\Cref{fig:partial-alignment-pictorial} illustrates how different devices may follow different branches of the function defined above, depending on local conditions. The figure shows that in each round, only one branch is executed per device—the other remains inactive (dashed). Some sub-expressions before branching still align across all devices (highlighted by red dashed arrows), but once branches diverge, exchange and aggregation occur strictly within the respective groups, with no cross-branch communication.


\subsection{Semantics} \label{ssec:opsem}

The meaning of \correction{\sysname}\ expressions is captured by two complementary semantic models. First, a big-step operational \emph{device semantics} defines how a single device computes during an individual round of execution. Second, a denotational \emph{network semantics} describes how the local computations performed by devices collectively determine the evolution of the entire network over time.

\begin{figure*}[t!]
\centerline{\framebox[\linewidth]{$\begin{array}{l}
	\textbf{Auxiliary definitions:} \\ \!\!\!
	\begin{array}{@{\;\;}ccl@{\hspace{44mm}}r}
		\senstate & \BNFcce & \envmap{\overline\aname}{\overline\anyvaluealt} \text{ where } \aname = \sname \text{ or } \xname & {\footnotesize \mbox{sensor state}}
		\\
		\vtree & \BNFcce & {\mkvt{}{\overline{\vtree}} \; \BNFmid} \; \mkvt{\anyvaluealt}{ {\overline\vtree} } &   {\footnotesize \mbox{value-tree}}
		\\
		\Trees & \BNFcce & \envmap{\overline{\deviceId}}{\overline{\vtree}} &   {\footnotesize \mbox{value-tree environment}}
		\\[5pt]
	\end{array} \\ \!\!\!
	\begin{array}{@{\;\;}l}
		{\piIof{i}{\mkvt{}{\vtree_1,\ldots,\vtree_n}} = \vtree_i} \qquad
		{\piIof{i}{\mkvt{\anyvaluealt}{\vtree_1,\ldots,\vtree_n}} = \vtree_i} \qquad
		{\piIof{i}{\envmap{\overline\deviceId}{\overline\vtree}} = \envmap{\overline\deviceId}{\piIof{i}{\overline\vtree}}}
		\\[2pt]
		{\proj{\envmap{\overline\deviceId}{\overline\vtree}}{\funvalue} = \bp{\envmap{\deviceId_i}{\vtree_i} ~\mid~ \vtree_i = \mkvt{\anyvaluealt}{\overline\vtree'}, \; \nameOf(\anyvaluealt(\deviceId_i)) = \nameOf(\funvalue)}}
  \;\;
		\begin{array}{l}
			\nameOf(\bname) = \bname \\
			\nameOf(\funKname \xname(\overline{\xname}) \{ \e \}) = \name
		\end{array} \\[10pt]
	\end{array} \\
	\hline\\[-10pt]
	\textbf{Evaluation rules:} \hspace{67mm} \boxed{\eqhlbsopsem{\deviceId}{\Trees}{\senstate}{\e}{\anyvaluealt}{\vtree}}
	\\
	\begin{array}{c}
		\nullsurfaceTyping{E-NVAL}{
			\!\!\! \eqhlbsopsem{\deviceId}{\Trees}{\senstate}{\anyvaluealt}{\anyvaluealt}{\mkvt{}{}} \!\!\!\!\!
		}
		\qquad
		\nullsurfaceTyping{E-LIT}{
			\!\!\! \eqhlbsopsem{\deviceId}{\Trees}{\senstate}{\lvalue}{\lvalue[]}{\mkvt{}{}} \!\!\!\!\!
		}
		\qquad
		\extension{
		\surfaceTyping{E-VAR}{
			\senstate(\xname) = \anyvaluealt
		}{
			\!\!\! \eqhlbsopsem{\deviceId}{\Trees}{\senstate}{\xname}{\anyvaluealt}{\mkvt{}{}} \!\!\!\!\!
		}}
		\skiptransition
		\surfaceTyping{E-VAL}{
			\begin{array}{l}
				\eqhlbsopsem{\deviceId}{\piI{1}(\Trees)}{\senstate}{\e_1}{\anyvaluealt_1}{\vtree_1} \\
				\eqhlbsopsem{\deviceId}{\piI{2}(\Trees)}{\senstate}{\applySubstitution{\e_2}{\substitution{\xname}{\anyvaluealt_1}}}{\anyvaluealt_2}{\vtree_2}
			\end{array}
		}{
			\eqhlbsopsem{\deviceId}{\Trees}{\senstate}{\valK \xname = \e_1; \e_2}{\anyvaluealt_2}{\mkvt{}{\vtree_1, \vtree_2}} \!\!\!\!\!
		}
			\skiptransition
		\surfaceTyping{E-APP}{
		\begin{array}{l}
			\eqhlbsopsem{\deviceId}{\piI{i+1}(\Trees)}{\senstate}{\e_i}{\anyvaluealt_i}{\vtree_i} \quad
			\text{for all}\; i \in 0, \ldots, n
			\\
			\eqhlauxopsem{\deviceId}{\piI{n+2}(\proj{\Trees}{\funvalue})}{\senstate}{\funvalue(\anyvaluealt_1, \ldots, \anyvaluealt_n)}{\anyvaluealt_{n+1}}{\vtree_{n+1}} \;
			\text{where } \funvalue = \anyvaluealt_0(\deviceId)
		\end{array} \!\!\!\!
		}{
			\eqhlbsopsem{\deviceId}{\Trees}{\senstate}{\e_0(\e_1,\ldots,\e_n)}{\anyvaluealt_{n+1}}{\mkvt{\funvalue[]}{\vtree_0,\ldots,\vtree_{n+1}}}
		}
		\\[20pt]
	\end{array} \\
	\hline\\[-10pt]
	\textbf{Auxiliary evaluation rules:} \hspace{45.5mm} \boxed{\eqhlauxopsem{\deviceId}{\Trees}{\senstate}{\funvalue(\overline\anyvaluealt)}{\anyvaluealt}{\vtree}} \\[-3mm]
	\quad
	\begin{array}{@{\!\!\!\!\!\!\!}c}
		\surfaceTyping{A-FUN}{
		\begin{array}{@{\!\!\!\!\!\!\!\!\!\!\!\!\!\!\!\!\!\!}l}
                   \\
                   \\
			\eqhlbsopsem{\deviceId}{\Trees}{\senstate}{\applySubstitution{\e}{\substitution{\xname}{\funKname \xname(\overline\xname)\{\e\}}, \substitution{\overline\xname}{\overline\anyvaluealt}}}{\anyvaluealt}{\vtree}
		\end{array} \!\!\!\!
		}{
             \begin{array}{c}
			\eqhlauxopsem{\deviceId}{\Trees}{\senstate}{\funKname \xname(\overline\xname)\{\e\}(\overline\anyvaluealt)}{\anyvaluealt}{\vtree}
               \\
               \\
               \end{array}
		}
		\\[-4mm]
		\extension{
		\surfaceTyping{A-SENS}{
			\senstate(\sname) = \anyvaluealt
		}{
			\eqhlauxopsem{\deviceId}{\Trees}{\senstate}{\sname()}{\anyvaluealt}{\mkvt{}{}}
		}}
		\qquad
		\nullsurfaceTyping{A-UID}{
			\eqhlauxopsem{\deviceId}{\Trees}{\senstate}{\uidK()}{\deviceId}{\mkvt{}{}}
		}
		\qquad
		\nullsurfaceTyping{A-SELF}{
			\eqhlauxopsem{\deviceId}{\Trees}{\senstate}{\selfK(\anyvaluealt)}{\anyvaluealt(\deviceId)}{\mkvt{}{}}
		}
            \skiptransition
            \surfaceTyping{A-XC}{
                \begin{array}{l}
                    \Trees = \envmap{\overline\deviceId}{\mkvt{\overline\anyvaluealt}{\ldots}} \quad
                    \anyvaluealt^\mathrm{nbr} = \anyvaluealt^\mathrm{init}[\envmap{\overline\deviceId}{\overline\anyvaluealt(\deviceId)}] \quad
                    \anyvaluealt^\mathrm{old} =
                    \begin{cases}
                        \proj{\anyvaluealt_k}{\overline\deviceId} & \text{if } \deviceId = \deviceId_k \text{ for a } k \\
                        \anyvaluealt^\mathrm{init} & \deviceId \text{ not in } \overline\deviceId
                    \end{cases}
                    \\
                    \eqhlbsopsem{\deviceId}{\piI{1}(\Trees)}{\senstate}{\anyvaluealt^\mathrm{fun}(\anyvaluealt^\mathrm{old}, \anyvaluealt^\mathrm{nbr})}{(\anyvaluealt^\mathrm{ret}, \anyvaluealt^\mathrm{send})}{\vtree}
                \end{array}
                \!\!\!\!\!
            }{
                \eqhlauxopsem{\deviceId}{\Trees}{\senstate}{\exchangeK(\anyvaluealt^\mathrm{init}, \anyvaluealt^\mathrm{fun})}{\anyvaluealt^\mathrm{ret}}{\mkvt{\anyvaluealt^\mathrm{send}}{\vtree}}
            }
            \skiptransition
		\surfaceTyping{A-FOLD}{
			\begin{array}{l}
				\Trees = \envmap{\deviceId_1}{\vtree_1}, \ldots, \envmap{\deviceId_n}{\vtree_n}
				\qquad
				\lvalue_0 = \anyvaluealt_3(\deviceId)
				\\
				\begin{cases}
					\eqhlbsopsem{\deviceId}{\emptyset}{\senstate}{\anyvaluealt_1(\lvalue_{i-1}, \anyvaluealt_2(\deviceId_i))}{\lvalue_i[]}{\vtree} &
					\text{if } \deviceId_i \neq \deviceId
					\\
					\lvalue_i = \lvalue_{i-1} & \text{otherwise}
				\end{cases}
				 \text{ for } i \in 1, \ldots, n
			\end{array}
			\!\!\!\!\!
		}{
			\eqhlauxopsem{\deviceId}{\Trees}{\senstate}{\foldK(\anyvaluealt_1, \anyvaluealt_2, \anyvaluealt_3)}{\lvalue_n[]}{\mkvt{}{}}
		}
		\qquad
		\cdots
	\end{array} \!\!\!\!
\end{array}$}}
\caption{Device (big-step) operational semantics of \correction{\sysname}} \label{fig:pfc:opsem-dev}
\end{figure*}

\subsubsection{Device-level semantics}

\Cref{fig:pfc:opsem-dev} defines the device-level semantics, formalised by judgement $\bsopsem{\deviceId}{\Trees}{\senstate}{\e}{\anyvaluealt}{\vtree}$, which states that ``expression $\e$ evaluates to \excvalue\ $\anyvaluealt$ and value-tree $\vtree$ on device $\deviceId$ with respect to sensor values $\senstate$ and value-tree environment $\Trees$'', where:
\begin{itemize}
    \item
    $\anyvaluealt$ is the final \emph{result} computed by evaluating $\e$;
    \item
    $\vtree$ is a value-tree, i.e., a representation of \excvalues\ encountered during evaluation of $\e$ as an ordered tree whose nodes store results from exchange expressions and function call stack frames (see \Cref{fig:pfc:opsem-dev}, top);
    \item
    $\Trees$ is a mapping from neighbouring device identifiers to their most recently produced value-trees, i.e., $\envmap{\deviceId_1}{\vtree_1}, \ldots, \envmap{\deviceId_n}{\vtree_n}$;
    \item
    $\senstate$ maps sensor names $\sname$ (and potentially free variables $\xname$ in open expressions\footnote{Note that defining semantics for open expressions is important to support inductive reasoning in theoretical proofs.}) to their respective \excvalues.
\end{itemize}
This semantics uses value-trees as communication structures. Each tree compactly represents all the data related to $\exchangeK$s generated during a round, while its shape reflects the execution's control flow—encoding alignment metadata via the nesting of function calls and subexpressions.
Before evaluation begins, every function expression $\funK \xname(\overline\xname)\{\e\}$ in the program is uniquely labelled with a name $\name$, \correction{that is used to define the function equality relation needed to ensure message alignment.} Thus, annotated functions are denoted as $\funKname \xname(\overline\xname){\e}$, and similarly for built-ins, with each carrying its own name $\bname$.

The syntax of value-trees and environments is summarised in \Cref{fig:pfc:opsem-dev} (top). The evaluation rules for judgement $\bsopsem{\deviceId}{\Trees}{\senstate}{\e}{\anyvalue}{\vtree}$ (\Cref{fig:pfc:opsem-dev}, middle) mostly follow standard constructs from functional languages, but are extended to incorporate alignment of neighbouring value-trees via the auxiliary projection functions $\piI{i}$.
Specifically, evaluating a subexpression $\e'$ uses a restricted environment $\Trees'$, obtained by applying $\piI{i}$ to extract the $i$-th subtree from each value-tree in $\Trees$. When $\piI{i}$ is applied to a tree $\vtree$, it returns its $i$-th child subtree; when applied to a full environment, it acts pointwise over all trees.
This alignment mechanism guarantees that $\Trees$ only includes subtrees corresponding to the same subexpression (in the same AST location and function call stack). In the case of function applications $\funvalue(\overline\anyvaluealt)$, alignment is further refined by reducing the environment to the subset $\proj{\Trees}{\funvalue}$—value-trees produced by prior evaluations of the same function, identified via its unique \emph{name} (as defined in \Cref{fig:pfc:opsem-dev} (top)).

The operational rules are as follows.
Rules \ruleNameSize{[E-NVAL]}, \ruleNameSize{[E-LIT]}, and \ruleNameSize{[E-VAR]} handle evaluation of constants and free variables, producing trivial (empty) value-trees and trivial results (the expressions themselves, or values in the sensor-state for \ruleNameSize{[E-VAR]}).
Rule \ruleNameSize{[E-VAL]} evaluates the first subexpression obtaining an nvalue $\anyvaluealt_1$, then substitutes it into the second subexpression; with evaluations relative to the two respective subtrees of $\Trees$.
Rule \ruleNameSize{[E-APP]} handles eager function applications: it evaluates the function expression and each argument using successive subtrees $\piIof{i+1}{\Trees}$. The application is then evaluated using the function-specific subtree $\piIof{n+2}{\proj{\Trees}{\funvalue}}$, ensuring alignment with earlier invocations of the same function. This application is delegated to auxiliary rules, defining how different  kinds of functions are applied.

Rule \ruleNameSize{[A-FUN]} substitutes the function’s arguments and name before evaluating its body.
Rules \ruleNameSize{[A-UID]}, \ruleNameSize{[A-SELF]}, and \ruleNameSize{[A-SENS]} handle built-ins directly using the device ID, device identity, or sensor values from $\senstate$, respectively.
Rule \ruleNameSize{[A-FOLD]} defines folding over \excvalues: the devices in $\Trees$ are enumerated in a consistent order, and a sequence of fold operations is applied using a binary function and per-device values, skipping the value for the current device, and using the third argument $\anyvaluealt_3$ as initial value.

The most notable rule, \ruleNameSize{[A-XC]}, governs the behaviour of exchange-expressions. As introduced in \Cref{ssec:communication}, this rule performs the key communication step. The notation $\evalue{\anyvaluealt^\mathrm{init}}{\overline\deviceId}{\overline\anyvaluealt(\deviceId)}$ represents the local exchange value, initialized with $\anyvaluealt^\mathrm{init}$ and updated with incoming messages from neighbours $\overline\deviceId$. For calculating the old value $\anyvaluealt^\mathrm{old}$, it distinguishes two cases:
\emph{(i)} if the device $\deviceId$ does not appear in $\Trees$, then the expression is being computed for the first time in the current round, so $\anyvaluealt^\mathrm{old}$ is set to the initial value;
\emph{(ii)} otherwise, if $\deviceId$ appears among the neighbours (i.e., $\deviceId = \deviceId_k$), then the previous round’s sent value $\anyvaluealt_k$ is recovered and its domain restricted to the current neighbourhood via $\proj{\anyvaluealt_k}{D}$, that corresponds to setting the value in $\anyvaluealt$ for any device $\deviceId'$ not in $D$ to the default value of $\anyvaluealt$.
Both the old value $\anyvaluealt^\mathrm{old}$ and the set of neighbour messages $\anyvaluealt^\mathrm{nbr}$ are passed to the update function $\anyvaluealt^\mathrm{fun}$. The first element of the result tuple is returned as the result of the exchange expression, while the second is stored in the root of the value-tree to be shared in the next round.
The semantics of other built-ins is standard and independent of $\Trees$, so these are omitted for brevity.

\begin{figure*}[t]
    \vspace{-10pt}
    \raisebox{-0.5\height}{\includegraphics[width=0.49\textwidth]{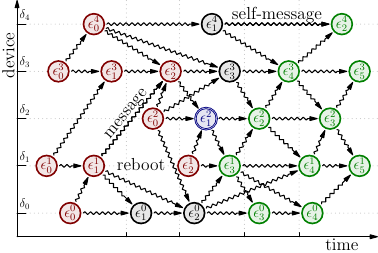}}
	\raisebox{-0.5\height}{\phantom{$\toSymK$} \hspace{-10pt}}
	\raisebox{-0.5\height}{\includegraphics[width=0.49\textwidth]{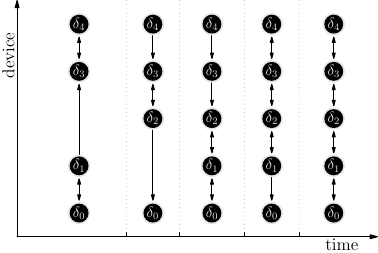}}

\caption{\textbf{Left:} An augmented event structure $\aEventS = \ap{\eventS,\neigh,\devof,\sensof}$
	where $\aEventS =
	\{\eventId^0_0,...,\eventId^0_4,
	\eventId^1_0,...,\eventId^1_5,
	\eventId^2_0,...,\eventId^2_4,
	\eventId^3_0,...,\eventId^3_3,
	\eventId^4_0,...,\eventId^4_5\}$  consists of 24 events such that $\devof(\eventId^i_j) =\deviceId_i$;
	it depicts events (circle nodes), messaging relations (curly arrows), devices $\deviceId_0,...,\deviceId_4$ (y-axis) and
	each circle node is labelled with the depicted event of $\aEventS$.
	Colours denote the causal relation w.r.t. the reference event $\eventId^2_1$ (doubly-circled, blue),
	partitioning events into causal past (red), causal future (green) and concurrent (non-ordered, in black).
	\textbf{Right:} A (possible) evolving network topology on devices which can give rise to $\aEventS$ (assuming some selected messages to be dropped).
} \label{fig:structure}\label{fig:evalue}
\end{figure*}

\subsubsection{Network-level semantics}

The behaviour of a network executing a program $\emain$ is modelled in terms of \emph{events}, which represent atomic rounds of execution performed by individual devices according to the device-level semantics. When considered collectively—spanning across both spatial dimensions (i.e., different devices) and temporal progression (i.e., the ordering of events)—these events define an execution trace of the program as if it were running on a unified ``aggregate machine'' \cite{BPV-COMPUTER2015}. Within this framework, communication between devices is captured via a \emph{messaging} relation $\neigh$, which governs how information is passed from one event to another.
To formalise this idea, we adopt the notion of \emph{augmented event structures}, which extend the classical \emph{event structures} from concurrency theory \cite{lamport:events} by enriching each event with contextual metadata: the associated device identifier and the local sensor state at the time of the event.

\begin{definition}[Augmented Event Structure] \label{def:augmentedES}
    An \emph{augmented event structure} ${\aEventS = \ap{\eventS,\neigh,\devof,\sensof}}$ is a tuple where:
    \begin{itemize}
        \item
        $\eventS$ is a countable set of \emph{events} $\eventId$,
        \item
        $\neigh \; \subseteq \eventS \times \eventS$ is a \emph{messaging} relation,
        \item
        $\devof : \eventS \to \deviceS$ is a mapping from events to the devices in which they occurred, and
        \item
        $\sensof: \eventS \to S$ maps each event $\eventId$ to a \emph{sensors status} $\senstate$ (as in the device-level semantics),
    \end{itemize}
    such that:
    \begin{itemize}
        \item $\eventId_1 = \eventId_2$ whenever $\devof (\eventId_1) = \devof(\eventId_2)$ and $\eventId_i \neigh \eventId$ for $i=1,2$  (i.e., precursors of events need to have occurred on distinct devices);
        \item there are no sequences $\eventId_1 \neigh \ldots \neigh \eventId_n \neigh \eventId_1$ (i.e., the $\neigh$ relation is acyclic);
        \item the set $X_\eventId = \bp{\eventId' \in \eventS \mid ~ \eventId' \neigh \ldots \neigh \eventId}$ of events that can reach $\eventId$ in $\neigh$ is finite for all $\eventId$ (i.e., $\neigh$ is well-founded and locally finite).
    \end{itemize}
\end{definition}

If $\eventId' \neigh \eventId$, we say that event $\eventId'$ is a \emph{\supplier} of event $\eventId$. We also refer to the transitive closure of $\neigh$ as the \emph{causality partial order} $< \; \subseteq \eventS \times \eventS$.

An illustrative example is given in \Cref{fig:structure} (top left), based on the dynamic network depicted in \Cref{fig:structure} (top right). The example showcases how the \emph{causality} relation segments the event space relative to a reference event $\eventId$, delineating a ``causal past'', ``causal future'', and a set of events concurrent with $\eventId$ (its ``present''). During execution, computations at $\eventId$ can rely on data produced in its past, and may in turn influence its future. The \emph{messaging relation} $\neigh$ defines when such influence can occur directly, i.e., when a message sent from one event can be consumed by another.

This model reflects the operational dynamics of distributed systems composed of interacting physical devices. The evolution of a single device corresponds to a chain of events linked by $\neigh$ and tagged with the same device ID, while inter-device communication is similarly represented by $\neigh$ links connecting events with different device tags. Importantly, this model abstracts away from assumptions such as global clocks, synchronized execution, or regular timing of events. Following prior literature~\cite{Mamei:2004a,DBLP:journals/corr/Lluch-LafuenteL16,Viroli:TOMACSselfstab}, we define a \emph{field of values} as a mapping from devices to values, representing a global snapshot of outputs produced by the most recent event on each device. The evolution of such fields over time forms a global structure called a \emph{space-time value}, assigning a value to each event in an augmented event structure.

\begin{definition}[Space-Time Value]\label{def:stvalue}
    Let  $\aEventS = \ap{\eventS,\neigh,\devof,\sensof}$ be an augmented event structure and $\setVS_\type$ be the domain of \excvalues\ of type $\type$.
    A \emph{space-time value} (in $\aEventS$ of type $\type$)  is a function mapping events to \excvalues\ $\dvalue : \eventS \to \setVS_\type$.
\end{definition}

Within this setting, evaluating a program $\emain$ on an augmented event structure $\aEventS$ induces a space-time value through iterative application of the device-level operational rules.

\begin{definition}[Program Evaluation on Event structures] \label{def:network:semantics}
    Let $\e$ be an \correction{\sysname}\ expression of type $\type$ given assumptions $\TtypEnv$.
    Let $\aEventS$ be an augmented event structure whose $\sensof$ includes values of the appropriate type for each sensor and free variable appearing in $\e$.
    
    Let $\vtree^{\aEventS}_{\e} : \eventS \to \TreeSet$ (where $\TreeSet$ is the set of all value-trees) and $\dvalue^{\aEventS}_{\e} : \eventS \to \setVS_{\type}$ be defined by induction on $\eventId$ in $\eventS$, so that $\eqhlbsopsem{\devof(\eventId)}{\Trees_{\eventId}}{\sensof(\eventId)}{\e}{\dvalue^{\aEventS}_{\e}(\eventId)}{\vtree^{\aEventS}_{\e}(\eventId)}$ where $\Trees_{\eventId} = \{\devof(\eventId') \mapsto \vtree^{\aEventS}_{\e}(\eventId') : \eventId' \neigh \eventId\}$. Then we say that $\dvalue^{\aEventS}_{\e} : \eventS \to \setVS_{\type}$ is the evaluation of expression $\e$ on $\aEventS$.
\end{definition}

It is important to observe that each event $\eventId$ is evaluated relative to a fixed value-tree environment $\Trees_\eventId$, which remains constant throughout the event’s execution.\footnote{This models a buffering abstraction on incoming messages, ensuring determinism within each round by assuming a consistent input snapshot.} Moreover, not all programs are guaranteed to produce a valid interpretation over an event structure $\aEventS$. Specifically, if execution of $\emain$ diverges in some event $\eventId$, the semantics becomes undefined for that structure. In real-world scenarios, this corresponds to a device failing to complete its computation: such an occurrence is indistinguishable from the perspective of the network from a device that has dropped out. Since non-terminating executions never emit output messages, they cannot contribute to the communication graph and thus do not constitute valid events in our model. Consequently, the semantics naturally excludes such pathological cases.

The semantics of a program $\emain$ can thus be interpreted from two complementary perspectives, local and global, echoing the micro/macro dichotomy common in self-organising systems \cite{SpatialIGI2013,regiment,kairos}.
From the \emph{local} (or micro) viewpoint, grounded in the device-level semantics, evaluating $\emain$ in a particular round $\eventId$ entails executing the expression on a device $\deviceId$, using the current sensor state $\senstate$ and the message environment $\Trees$—which encodes inputs from neighbouring events—producing an output value $\anyvaluealt$ and an associated message structure $\vtree$ to be disseminated.
From the \emph{global} (or macro) viewpoint, as formalised by the network-level semantics, the program $\emain$ acts as a partial functor over augmented event structures. Given an input $\aEventS = \ap{\eventS, \neigh, \devof, \sensof}$,\footnote{The functor is partial because the interpretation fails if the program diverges on any event in $\aEventS$.} the semantics yields a space-time value $\dvalue$ that encapsulates the outputs of the program across the full range of events in both space and time.


\section{XC Design Decision and Properties} \label{sec:design-and-properties}

\subsection{Design Decisions} \label{ssec:design}

\XC\ has been designed to
 encourage developers to program by defining and composing  functional blocks
 that are inherently resilient to failures. Namely, \XC\ enjoys the following three overall characteristics.

\subsubsection{Collective Viewpoint}
The $\exchangeK$ mechanism allows programmers to specify a logic to {\em collectively} operate over the neighbours' messages, expressing the behaviour homogeneously through point-wise operations and \lstinline|nfold|, without making assumptions on neighbours number or identity;  while the round-based execution model supports progress
 and incorporation of local changes (ultimately propagating from neighbourhoods to neighbourhoods up to the global system). 
Therefore, every behavior
 can be described by a function
 which encapsulates both the processing
 the communication necessary to achieve a coordinated result.

\subsubsection{Automatic failure absorption}
If a device fails or a message gets lost,  $\exchangeK$ handles the failure transparently from programmers: the device may not show up among the neighbours of some device in the next firing. 
Therefore,  \XC\ supports writing programs that are \emph{inherently} fault tolerant and resilient with respect to
 devices dynamically joining and leaving (transparently from programmers) the set of neighbours (e.g., because they physically change location or reboot).
\correction{That said, it is important to note that \XC, being a Turing-complete language, does not by itself enforce fault-tolerant behaviour. Programmers can still express non-resilient logic, and in fact such behaviour is unavoidable in a language of this expressive power. However, such non-resilient programs are typically non-idiomatic. Stronger guarantees can instead be established for well-defined, idiomatic subsets of the language, as discussed in \Cref{ssec:properties}.}

\subsubsection{Compositionality}
The interaction, governed by alignment, between $\exchangeK$ on the one hand and functional abstraction/composition and branching on the other, supports the compositional programming of resilient collective behaviour. Namely,
 functions expressing resilient blocks of collective behaviour (cf. gradients in \Cref{exmp:distance}) can be defined
 and
 composed together to build more complex blocks, while retaining properties like fault tolerance and self-stabilisation (see \Cref{ssec:properties}).

\subsection{Properties} \label{ssec:properties}

In this section we list the main properties of \XC. The following properties were demonstrated in~\cite{Audrito-et-al:JSS-2024} for a version of \XC\ that presents a simpler version of $\exchangeK$, extending the proofs to the version of $\exchangeK$ presented in \Cref{sec:fromLCtoXC} (which was introduced in~\cite{Audrito-et-al:COORDINATION-2024}) is  is straightforward. 
\begin{itemize}
    \item 
    Type soundness: well-typed program are such that any firing can not \correction{get} stuck.
    \item 
    Self stabilisation: programs written in a \correction{given idiomatic subset} of the language enjoy eventual recovery and stabilisation after transient changes, i.e., if the neighbouring relation and the values of the sensor stabilises than the value produced by any firing eventually stabilises.
\end{itemize}
Moreover, in~\cite{Audrito-et-al:JSS-2024} it is also pointed out that, since  every \FC{} program can be encoded within \XC{}, then the following results proved in the literature  for subsets of the \FC\ language:
\begin{itemize}
    \item 
independence of the results from the density of devices~\cite{BVPD-TAAS2017}, 
\item 
real-time error guarantees~\cite{a:rtssgradient}, 
\item 
ability to express all physically consistent computations (space-time universality)~\cite{a:fcuniversality}, and 
\item 
ability to monitor spatial/temporal logic formulas~\cite{DBLP:journals/jss/AudritoCDSV21,DBLP:journals/jss/AudritoDSTV22},
\end{itemize}
hold also for the corresponding subsets of the \XC\ language.

\section{FCPP: a C++ Library and Toolchain for the eXchange Calculus} \label{sec:fcpp}

FCCP, the C++ library and toochain  for \XC{} developed at the Department of Computer science of the University of Turin,
 is designed for \emph{(i)} efficiency, and \emph{(ii)} custom architectures.
For \emph{(i)}, FCPP relies on C++'s compile-time optimisation and execution on the bare metal.
Neighbouring values are implemented using \lstinline|vector<T>| 
from C++ STL (this solution  is more efficient than hash maps for linear folding and point-wise operations).
Communication relies \correction{on} message serialization and passing to the network driver
(which, for low level devices this is usually a device specific API).
For \emph{(ii)}, FCPP  exploits C/C++ compilers available for custom architectures and aims to simplify the deployment by minimising the amount of dependencies. E.g., by including into the implementation
 serialisation header, compile-time type inspection utilities, multi-type valued maps, option types,
quaternions, tagged tuples, etc.
This FCPP implementation has been tested and used on different back-ends which do not have external dependencies, including:
\begin{itemize}
	\item batch simulations of distributed networks, producing plots summarising the network behaviour across several runs {\cite{FCPP-ACSOS-2020};}
	\item graphical and interactive 3D simulation of a distributed network through OpenGL {\cite{DBLP:conf/coordination/AudritoRT22};}
	\item processing of \XC{} algorithms on large graph-based data in HPC {\cite{DBLP:conf/isola/AudritoDT22};}
	\item deployment on microcontroller architectures with either Contiki OS or MIOSIX {\cite{DBLP:journals/percom/TestaADT22,DBLP:journals/tpds/AudritoTF23}, also interoperable with Android phones (work in progress).}
\end{itemize}

\corrstart
The following snippet of code, reporting the implementation of XC functions $distanceEstimate$ and $distanceTo$ from \Cref{sec:communication}, gives a flavor of the use of FCPP in a C++ program.

\begin{lstlisting}[language=fcpp,mathescape]
FUN real_t distanceEstimate(ARGS, nvalue<real_t> n) { CODE
    return min_hood(CALL, n + node.nbr_dist(), INF);
}

FUN real_t distanceTo(ARGS, bool isSource) { CODE
  return exchange(CALL, INF, [&](nvalue<real_t> n) {
    return mux(isSource, 0, distanceEstimate(CALL, n));
  }  
}
\end{lstlisting}

Thanks to macros (like \lstinline[language=fcpp]{FUN, ARGS, CODE, CALL}), the FCPP library offers XC as an internal DSL (Domain Specific Language), i.e., the developer can mix ``XC code'' and C++ in the same program.
The basic operators and builtin functions of XC, (like \lstinline[language=fcpp]{min_hood, exchange, mux}), are defined as C++ functions, that exploit templates for polymorphism.
Despite revealing some unavoidable characteristics of C++ that differ from the formal definition of XC (e.g., explicit types, \lstinline[language=fcpp]{return} statements, etc.), the code in the snippet looks very close to the corresponding XC code shown in  \Cref{sec:communication}.

\corrend

\section{FCPP at Work} \label{sec:fcpp-at-work}

In this section we show how, thanks to some projects where we are involved, we have already put XC ``on wheels'' on board of autonomuos robots, and how we are working hard to make it ``fly'' on board of drones.

\paragraph{RoboNG} The RoboNG project\footnote{\url{https://ecs-nodes.eu/en/1-aerospace-and-sustainable-mobility/progetti-imprese/robong}} is funded by the European Union - NextGenerationEU within the NODES ecosystem. It started at the end of 2023 and lasted 14 months, until February 2025, involving two main industrial partners, and the Dipartimento di Informatica of Università di Torino as an additional academic partner.

The use case addressed within the project was supplied by SAGAT, the company that manages Torino Airport, the international airport serving the city of Turin (Italy). 
SAGAT proposed as a use case a relevant problem faced by airports, namely the timely detection and removal of FODs (Foreign Object Debris). FODs are small objects that can clutter the parking and \correction{taxiing} areas used by \correction{airplanes}, with potentially disastrous effects, including fatalities\footnote{\url{https://www.fodcontrol.com/what-is-fod/}}.
Since continuously inspecting the relevant areas for FOD detection is a repetitive, costly, and error-prone task when performed by humans, the RoboNG project has defined, as its main goal, the prototype implementation of a system able to perform FOD detection in a (semi) autonomous way.

The proposed solution involves a team of Autonomous Mobile Robots (AMRs) in charge of covering the relevant areas while exploiting on-board sensors to detect potential FODs, and send alerts to human operators. 
In order to be able to navigate the environment, perceive it, and communicate with the operator each robot is equipped with:
\begin{itemize}
    \item high-resolution camera to observe the area immediately in front of the robot
    \item Computer Vision (CV) pipeline, based on AI (executed on an on-board computer) for the detection of FODs
    \item autonomous navigation software (supporting obstacle and collision avoidance), powered by a Lidar and/or an additional camera
    \item WiFi connection to a local wireless network supplied by a router
\end{itemize}
Additionally, and more relevantly for our scientific interests, robots must be able to host on-board a FCPP application that enables communication and coordination among robots. As the main {\em robot brain}, the FCPP application gets information from the CV pipeline, and accesses the local robot to get telemetry and send control commands. 

\begin{figure}[t]
\centering
  \includegraphics[height=1.8in]{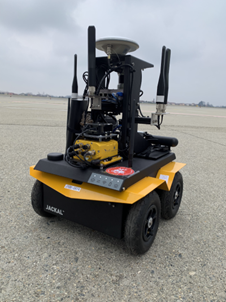}
  \hspace*{0.5in}
  \includegraphics[height=1.8in]{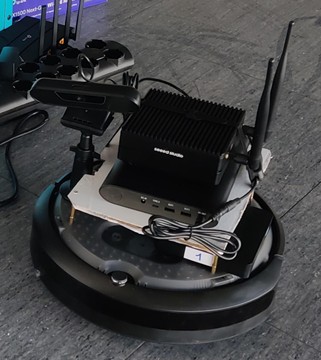}
  \caption{The Jackal (left) and Create3 (right) Robots equipped for RoboNG.}\label{fig:jackal}
\end{figure}

Figure~\ref{fig:jackal} shows the two robots employed for the prototype system implementation. The Jackal, produced by Clearpath Robotics, is a powerful small-size rover for outdoor exploration. The Create3,  produced by iRobot, is a basic platform for research, usable mainly indoor.
In the RoboNG final demo at Torino Airport, the two robots have cooperated to cover a small area where they have successfully detected and reported a few FODs. 
The distributed covering algorithm was implemented as an XC program running on the on-board FCPP-based application.
\emph{In simulations} with the Gazebo simulator, the XC program has been shown to successfully coordinate up to eight Jackal robots for the FOD detection task.

\paragraph{RoboApp} The RoboApp project\footnote{\url{https://ecs-nodes.eu/en/1-aerospace-and-sustainable-mobility/progetti-accademici/roboapp}}, is funded by the European Union - NextGenerationEU within the NODES ecosystem. It started at the beginning of 2024 and is almost ending at the time of writing (duration of 14 months). The partners are Dipartimento di Economia e Statistica, and Dipartimento di Informatica of Università di Torino, which have outsourced some parts of the work to small external enterprises.

SAGAT proposed another compelling use case for the project, involving the boarding gates instead of outdoor areas. In particular, it can happen that, due to several reasons, long queues of people form at the gates, overflowing out of the serpentine barriers into the transit areas. Such queues should be promptly managed by human operators, since they get in the way of people passing through, but it is often the case that queues form while no one is (yet) attending the gate.  
The RoboAPP project has defined as its main goal, the prototype implementation of a system able to detect overflowing queues at the gates, and warn human operators when needed.

The proposed solution has some similarities with the one devised for RoboNG, but presents key differences that make it a new research challenge of itself.
We consider again a team of Autonomous Mobile Robots (AMRs) in charge of monitoring the gates, equipped with sensors and software for navigating and observing the environment with a camera, as well as with an FCPP application for distributed control and coordination.
When the schedule (which is currently fixed) requires that a gate is monitored, the team executes a distributed XC algorithm to choose two (or more) robots that will jointly execute the task. The chosen robots navigate from the parking areas (where they gather when idle) to predefined locations close to the gate, where they can start their observations.
Up to this point the solution does not involve anything new with respect to RoboNG. The novelties are indeed in the way the robots cooperate to observe the scene, and in the steps of the decision making process for issuing an alert to the human operator.

\begin{figure}[t]
\centering
  \includegraphics[scale=0.25]{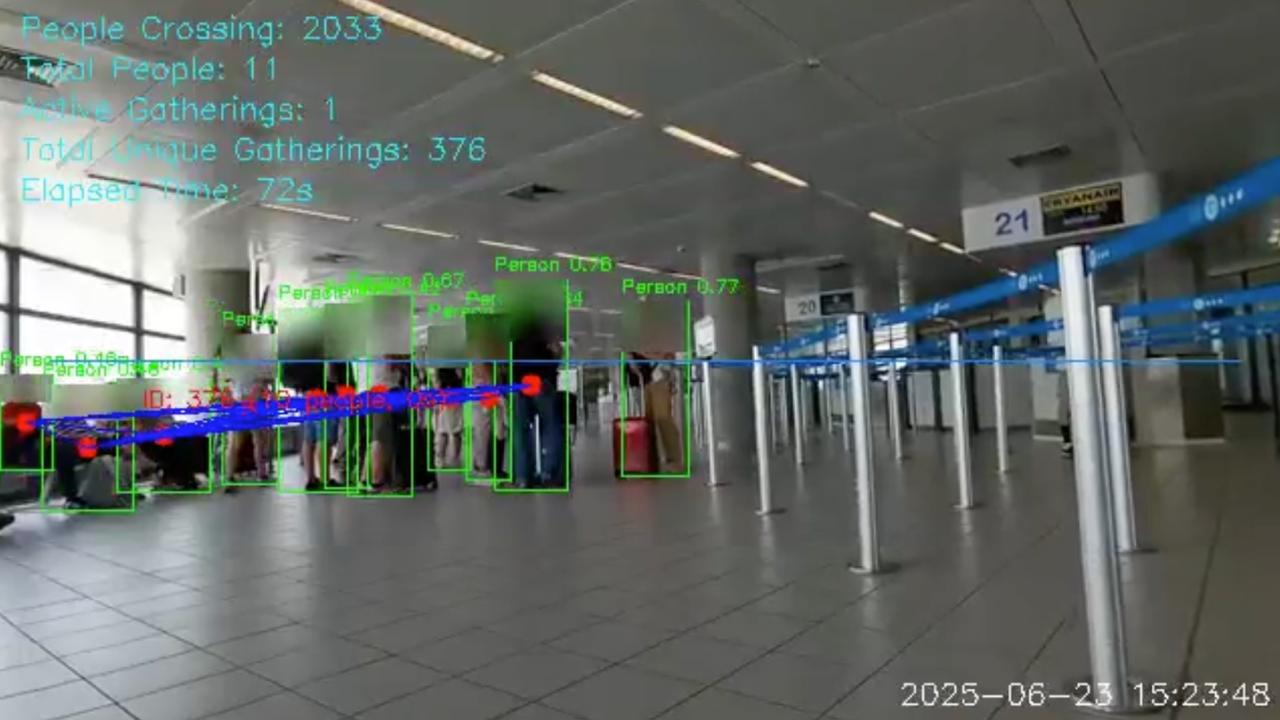}
  \caption{Frame processed by the RoboAPP CV pipeline.}\label{fig:queue}
\end{figure}

First of all, detecting a queue near a gate is a much more challenging computer-vision task than detecting a FOD in a mostly uniform and empty area, and the CV pipeline (executed on on-board Jetson Nano computers) is therefore correspondingly more complex. In particular, the AI model is required to detect individual people with their bounding boxes; then, further processing steps must deal with distances and time to detect stable gatherings of people, that may indicate queues (see Figure~\ref{fig:queue}).
More importantly, each robot perceives the scene from its own point of view, and may draw different conclusions about the presence of an overflowing queue. We exploit again an FCPP application running onboard, that implements a distributed XC algorithm to reach a consensus among the robots (in the simplest case, a distributed majority voting).

However, even the consensus among multiple points of view is sometimes unreliable for the task at hand, since what we need to recognize involves spatial, temporal and common sense knowledge that is not available to CV models. For example, we would like to avoid sending alerts in situations where there's actually a queue, but it does not obstruct the passage of other people (e.g., because they are within the serpentine barrier, or because they are lining outside of the serpentine in an appropriate way).
We have therefore added an additional component to the system, namely an AI Agents \cite{wang2024survey} application running on a PC on the local network, based on a multi-modal Small Language Model (SLM). The AI application is invoked to make a final decision when the team of robots agrees that there is {\em likely} a critical situation to notify to the human operators.
This architecture is an example of how Aggregate Programming and AI Agents may cooperate to improve the efficiency and accuracy of a system.

\paragraph{AgriTech} We partecipate to the Agritech National Research Center for Technology in Agriculture, Spoke 6\footnote{\url{https://agritechcenter.it/}}, which is funded by the European Union - NextGenerationEU. It started at the end of 2022 and is almost ending at the time of writing (duration of 3 years). 
Among the many partners, we have cooperated in particular with the Dipartimento di Scienze Agrarie, Forestali e Alimentari (DISAFA) of Università di Torino within the Task 6.1.3 \emph{Case studies and indicators-based evaluation of new management models for different agricultural scenarios}.

The part of the task relevant for our scientific interests, is the study of the employment of swarms of drones in agriculture. Guided by the domain experts at DISAFA, we have implemented several simulations of swarms of autonomous drones, controlled and coordinated with AP algorithms implemented in XC. More specifically, we have addressed the following problem: given an autonomous vehicle (either aerial or ground), named \emph{worker}, in charge of following a pre-determined trajectory across a crops area to perform its task (e.g., weeding, or
trimming), devise algorithms to coordinate a swarm of drones that must accompany the worker along its trajectory providing surveillance of the surrounding areas, for safety purposes. 
We have explored the following variants:
\begin{itemize}
    \item single worker and multi-worker: the problem becomes much more interesting (from the AP point of view) when instead of a single worker, there are several workers, each with its associated swarm, and individual drones can move from one swarm to another one to optimize monitoring
    \item lightweight simulation embedded in FCPP (where physics are simulated at a very abstract level, and drones are represented as spheres), and realistic simulation with Gazebo (where physics are realistic, and drones have a 3D model and can even execute their original on-board control firmware as software-in-the-loop)
\end{itemize}

\begin{figure}[t]
\centering
  \includegraphics[scale=0.5]{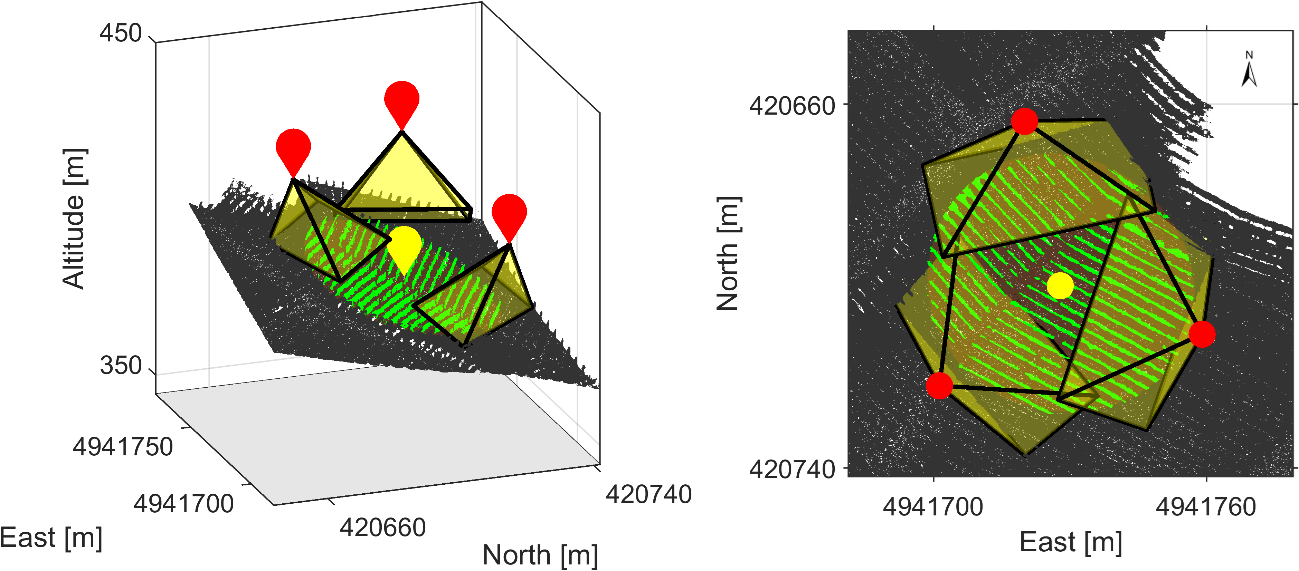}
  \caption{A UGV (yellow) and a swarm of three drones (red) in a vineyard (from \cite{nitti2025drone}.}\label{fig:vineyard}
\end{figure}

In \cite{nitti2025drone}, we have reported the results of a cooperation with DISAFA, where we have experimentally studied the effectiveness of the implementation based on a single worker and lightweight simulation in providing surveillance for a UGV in a vineyard, taking into account the detailed 3D map of the terrain and the field-of-view of the cameras mounted on the drones (Figure \ref{fig:vineyard}).

While the work within the Agritech project has been limited to simulations, we have recently started a six-months self-funded project named DRONEX, where we are deploying FCPP on a swarm of nano-drones, namely Crazyflies (CF) by Bitcraze. 
The small size of CFs imposes severe limitations to the computational power (and energy consumption) of the boards that we can mount on the drones to run FCPP and other software for the autonomous control.
We have chosen the Luckfox Pico mini boards that, despite their size, provide an ARM Cortex-A7 processor that can run an embedded Linux, and even a Neural Processing Unit (NPU) that we could exploit for on-board AI.
The very low requirements of FCPP in terms of CPU and, especially, of memory allowed us to deploy it and run it successfully on such small boards. 

\correction{
\section{Related Work} \label{sec:relwork}
When developing collective adaptive systems—such as robotic swarms or Wireless Sensor Networks (WSNs)—it is advantageous to program the system at a global level, abstracting away from low-level heterogeneity among computational nodes and from explicit communication mechanisms. This goal can be addressed through collective-oriented approaches to Distributed Systems Engineering (DSE), such as methods for Collective Adaptive Systems (CAS), and \emph{macro-programming}.
For a comprehensive and up-to-date survey of software engineering methodologies for collective systems, the reader is referred to~\cite{caadpstv:tosem25}.

The notion of CAS refers to systems composed of large numbers of interacting components that cooperate on complex tasks without any form of centralized control, and that operate in dynamic environments~\cite{DBLP:journals/corr/abs-1108-5643,DBLP:conf/saso/CabriH14,DBLP:journals/sttt/NicolaJW20}.
A significant portion of the languages proposed for the engineering of CASs concentrates on the specification and analysis of so-called \emph{ensembles}, that is, dynamic collections of devices characterized by rich interaction patterns and situated in complex environments~\cite{DBLP:conf/fmco/WirsingHTZ11}.
A representative example is \emph{CARMA}~\cite{DBLP:conf/sfm/LoretiH16}, which is based on a stochastic process algebra.
With respect to ensemble programming, the ASCENS project introduced \emph{SCEL}~\cite{DBLP:journals/taas/NicolaLPT14}, a process-algebraic abstract language for describing the behaviour of individual components as well as of their aggregations. SCEL relies on \emph{attribute-based communication}~\cite{DBLP:journals/scp/AlrahmanNL20} and supports specifications that are parametric with respect to knowledge repositories and adaptation policies.

Another relevant approach to collective-oriented DSE is \emph{macro-programming}~\cite{DBLP:journals/csur/Casadei23}. These approaches provide collective-level abstractions in which program constructs are defined over the entire ensemble of devices, effectively treating the distributed system itself as the execution target.
Representative proposals based on the macro-programming approach include:
the \emph{Buzz} swarm programming DSL~\cite{DBLP:journals/software/PinciroliB16}, which introduces \emph{swarms} as first-class entities;
\emph{AErlang}~\cite{DBLP:journals/scp/NicolaDIT18}, a SCEL-based DSL that exploits attribute-based communication;
\emph{Chronus}~\cite{wadaa2010chronus}, a spacetime-oriented programming language for wireless sensor networks in which programs specify data queries over ``space-time slices'';
and \emph{Aggregate Programming}~\cite{DBLP:journals/computer/BealPV15}, which relies on the notion of \emph{computational fields}~\cite{DBLP:conf/esaw/MameiZL02}.
%
}

\section{Conclusion and Future Work} \label{sec:conclusion}

We have provided an introduction (tailored for researchers who, like Stefano Berardi, have research interests in lambda calculus and type theory) to the AP paradigm and its foundational model XC, together with a briefly overview of some industrial case studies 
involving the FCCP C++ library and  internal DSL incarnation of XC. Ongoing research is addressing both foundational aspects, like:
\begin{itemize}
    \item adding to XC declarative mechanisms  for exploiting devices heterogeneity (currently this can only be done by relying on $\ifK$-expressions); 
    \item 
    further investigating real-time guarantees; and
    \item 
    designing more expressive XC-\correction{monitorable} spatial-temporal logics; 
\end{itemize}
and practical  aspects, for instance:
\begin{itemize}
    \item a refactoring of FCPP is planned; 
    \item 
    a Kotlin library and internal DSL incarnation  of XC is currently under development (\url{https://collektive.github.io/}) by a group of researchers at the Department of Computer Science and Engineering of
the University of Bologna); and
\item 
further validation of the approach by real-world use-cases is planned.
\end{itemize}

\subsection*{Acknowledgements}
This publication is part of the project NODES,
which has received funding from the MUR – M4C2 1.5 of PNRR funded by
the European Union - NextGenerationEU (Grant agreement no.
ECS00000036). This  study has been supported by the Italian PRIN project ``CommonWears'' (2020HCWWLP) and
was carried out within the Agritech National Research Center and received funding from the European Union Next-GenerationEU (PIANO NAZIONALE DI RIPRESA E RESILIENZA (PNRR) – MISSIONE 4 COMPONENTE 2, INVESTIMENTO 1.4 – D.D. 1032 17/06/2022, CN00000022). This manuscript reflects only the authors’ views and opinions, neither the European Union nor the European Commission can be considered responsible for them.

\bibliographystyle{eptcs}
\bibliography{long}

\begin{thebibliography}{10}
\providecommand{\bibitemdeclare}[2]{}
\providecommand{\surnamestart}{}
\providecommand{\surnameend}{}
\providecommand{\urlprefix}{Available at }
\providecommand{\url}[1]{\texttt{#1}}
\providecommand{\href}[2]{\texttt{#2}}
\providecommand{\urlalt}[2]{\href{#1}{#2}}
\providecommand{\doi}[1]{doi:\urlalt{https://doi.org/#1}{#1}}
\providecommand{\eprint}[1]{arXiv:\urlalt{https://arxiv.org/abs/#1}{#1}}
\providecommand{\bibinfo}[2]{#2}

\bibitemdeclare{article}{DBLP:journals/scp/AlrahmanNL20}
\bibitem{DBLP:journals/scp/AlrahmanNL20}
\bibinfo{author}{Yehia~Abd \surnamestart Alrahman\surnameend},
  \bibinfo{author}{Rocco~De \surnamestart Nicola\surnameend} \&
  \bibinfo{author}{Michele \surnamestart Loreti\surnameend}
  (\bibinfo{year}{2020}): \emph{\bibinfo{title}{Programming interactions in
  collective adaptive systems by relying on attribute-based communication}}.
\newblock {\slshape \bibinfo{journal}{Sci. Comput. Program.}}
  \bibinfo{volume}{192}, p. \bibinfo{pages}{102428},
  \doi{10.1016/J.SCICO.2020.102428}.

\bibitemdeclare{inproceedings}{FCPP-ACSOS-2020}
\bibitem{FCPP-ACSOS-2020}
\bibinfo{author}{Giorgio \surnamestart Audrito\surnameend}
  (\bibinfo{year}{2020}): \emph{\bibinfo{title}{{FCPP:} an efficient and
  extensible field calculus framework}}.
\newblock In: {\slshape \bibinfo{booktitle}{Proceedings of the 1st
  International Conference on Autonomic Computing and Self-Organizing Systems,
  {ACSOS}}}, \bibinfo{publisher}{IEEE Computer Society}, pp.
  \bibinfo{pages}{153--159}, \doi{10.1109/ACSOS49614.2020.00037}.

\bibitemdeclare{article}{DBLP:journals/corr/abs-1910-02874}
\bibitem{DBLP:journals/corr/abs-1910-02874}
\bibinfo{author}{Giorgio \surnamestart Audrito\surnameend},
  \bibinfo{author}{Jacob \surnamestart Beal\surnameend},
  \bibinfo{author}{Ferruccio \surnamestart Damiani\surnameend},
  \bibinfo{author}{Danilo \surnamestart Pianini\surnameend} \&
  \bibinfo{author}{Mirko \surnamestart Viroli\surnameend}
  (\bibinfo{year}{2020}): \emph{\bibinfo{title}{Field-based Coordination with
  the Share Operator}}.
\newblock {\slshape \bibinfo{journal}{Logical Methods in Computer Science}}
  \bibinfo{volume}{16(4)}, \doi{10.23638/LMCS-16(4:1)2020}.

\bibitemdeclare{inproceedings}{a:fcuniversality}
\bibitem{a:fcuniversality}
\bibinfo{author}{Giorgio \surnamestart Audrito\surnameend},
  \bibinfo{author}{Jacob \surnamestart Beal\surnameend},
  \bibinfo{author}{Ferruccio \surnamestart Damiani\surnameend} \&
  \bibinfo{author}{Mirko \surnamestart Viroli\surnameend}
  (\bibinfo{year}{2018}): \emph{\bibinfo{title}{Space-Time Universality of
  Field Calculus}}.
\newblock In: {\slshape \bibinfo{booktitle}{Coordination Models and
  Languages}}, {\slshape \bibinfo{series}{Lecture Notes in Computer Science}}
  \bibinfo{volume}{10852}, \bibinfo{publisher}{Springer}, pp.
  \bibinfo{pages}{1--20}, \doi{10.1007/978-3-319-92408-3_1}.

\bibitemdeclare{inproceedings}{Audrito-et-al:COORDINATION-2024}
\bibitem{Audrito-et-al:COORDINATION-2024}
\bibinfo{author}{Giorgio \surnamestart Audrito\surnameend},
  \bibinfo{author}{Daniele \surnamestart Bortoluzzi\surnameend},
  \bibinfo{author}{Ferruccio \surnamestart Damiani\surnameend},
  \bibinfo{author}{Giordano \surnamestart Scarso\surnameend} \&
  \bibinfo{author}{Gianluca \surnamestart Torta\surnameend}
  (\bibinfo{year}{2024}): \emph{\bibinfo{title}{An Enhanced Exchange Operator
  for XC}}.
\newblock In \bibinfo{editor}{Ilaria \surnamestart Castellani\surnameend} \&
  \bibinfo{editor}{Francesco \surnamestart Tiezzi\surnameend}, editors:
  {\slshape \bibinfo{booktitle}{Coordination Models and Languages}},
  \bibinfo{publisher}{Springer Nature Switzerland}, \bibinfo{address}{Cham},
  pp. \bibinfo{pages}{129--145}, \doi{10.1007/978-3-031-62697-5_8}.

\bibitemdeclare{inproceedings}{DBLP:conf/ecoop/AudritoCDSV22}
\bibitem{DBLP:conf/ecoop/AudritoCDSV22}
\bibinfo{author}{Giorgio \surnamestart Audrito\surnameend},
  \bibinfo{author}{Roberto \surnamestart Casadei\surnameend},
  \bibinfo{author}{Ferruccio \surnamestart Damiani\surnameend},
  \bibinfo{author}{Guido \surnamestart Salvaneschi\surnameend} \&
  \bibinfo{author}{Mirko \surnamestart Viroli\surnameend}
  (\bibinfo{year}{2022}): \emph{\bibinfo{title}{Functional Programming for
  Distributed Systems with {XC}}}.
\newblock In: {\slshape \bibinfo{booktitle}{36th European Conference on
  Object-Oriented Programming, {ECOOP} 2022}}, {\slshape
  \bibinfo{series}{LIPIcs}} \bibinfo{volume}{222}, \bibinfo{publisher}{Schloss
  Dagstuhl}, pp. \bibinfo{pages}{20:1--20:28},
  \doi{10.4230/LIPIcs.ECOOP.2022.20}.

\bibitemdeclare{article}{Audrito-et-al:JSS-2024}
\bibitem{Audrito-et-al:JSS-2024}
\bibinfo{author}{Giorgio \surnamestart Audrito\surnameend},
  \bibinfo{author}{Roberto \surnamestart Casadei\surnameend},
  \bibinfo{author}{Ferruccio \surnamestart Damiani\surnameend},
  \bibinfo{author}{Guido \surnamestart Salvaneschi\surnameend} \&
  \bibinfo{author}{Mirko \surnamestart Viroli\surnameend}
  (\bibinfo{year}{2024}): \emph{\bibinfo{title}{The eXchange Calculus (XC): A
  functional programming language design for distributed collective systems}}.
\newblock {\slshape \bibinfo{journal}{Journal of Systems and Software}}
  \bibinfo{volume}{210}, p. \bibinfo{pages}{111976},
  \doi{10.1016/j.jss.2024.111976}.

\bibitemdeclare{article}{DBLP:journals/jss/AudritoCDSV21}
\bibitem{DBLP:journals/jss/AudritoCDSV21}
\bibinfo{author}{Giorgio \surnamestart Audrito\surnameend},
  \bibinfo{author}{Roberto \surnamestart Casadei\surnameend},
  \bibinfo{author}{Ferruccio \surnamestart Damiani\surnameend},
  \bibinfo{author}{Volker \surnamestart Stolz\surnameend} \&
  \bibinfo{author}{Mirko \surnamestart Viroli\surnameend}
  (\bibinfo{year}{2021}): \emph{\bibinfo{title}{Adaptive distributed monitors
  of spatial properties for cyber-physical systems}}.
\newblock {\slshape \bibinfo{journal}{J. Syst. Softw.}} \bibinfo{volume}{175},
  p. \bibinfo{pages}{110908}, \doi{10.1016/j.jss.2021.110908}.

\bibitemdeclare{inproceedings}{audrito2017ULT}
\bibitem{audrito2017ULT}
\bibinfo{author}{Giorgio \surnamestart Audrito\surnameend},
  \bibinfo{author}{Roberto \surnamestart Casadei\surnameend},
  \bibinfo{author}{Ferruccio \surnamestart Damiani\surnameend} \&
  \bibinfo{author}{Mirko \surnamestart Viroli\surnameend}
  (\bibinfo{year}{2017}): \emph{\bibinfo{title}{Compositional Blocks for
  Optimal Self-Healing Gradients}}.
\newblock In: {\slshape \bibinfo{booktitle}{Self-Adaptive and Self-Organizing
  Systems (SASO), 2017}}, \bibinfo{organization}{IEEE},
  \bibinfo{publisher}{{IEEE} Computer Society}, pp. \bibinfo{pages}{91--100},
  \doi{10.1109/SASO.2017.18}.

\bibitemdeclare{article}{DBLP:journals/lmcs/AudritoCDV23}
\bibitem{DBLP:journals/lmcs/AudritoCDV23}
\bibinfo{author}{Giorgio \surnamestart Audrito\surnameend},
  \bibinfo{author}{Roberto \surnamestart Casadei\surnameend},
  \bibinfo{author}{Ferruccio \surnamestart Damiani\surnameend} \&
  \bibinfo{author}{Mirko \surnamestart Viroli\surnameend}
  (\bibinfo{year}{2023}): \emph{\bibinfo{title}{Computation Against a
  Neighbour: Addressing Large-Scale Distribution and Adaptivity with Functional
  Programming and {Scala}}}.
\newblock {\slshape \bibinfo{journal}{Log. Methods Comput. Sci.}}
  \bibinfo{volume}{19}(\bibinfo{number}{1}), \doi{10.46298/lmcs-19(1:6)2023}.

\bibitemdeclare{article}{DBLP:journals/jss/AudritoDSTV22}
\bibitem{DBLP:journals/jss/AudritoDSTV22}
\bibinfo{author}{Giorgio \surnamestart Audrito\surnameend},
  \bibinfo{author}{Ferruccio \surnamestart Damiani\surnameend},
  \bibinfo{author}{Volker \surnamestart Stolz\surnameend},
  \bibinfo{author}{Gianluca \surnamestart Torta\surnameend} \&
  \bibinfo{author}{Mirko \surnamestart Viroli\surnameend}
  (\bibinfo{year}{2022}): \emph{\bibinfo{title}{Distributed runtime
  verification by {past-CTL} and the field calculus}}.
\newblock {\slshape \bibinfo{journal}{J. Syst. Softw.}} \bibinfo{volume}{187},
  p. \bibinfo{pages}{111251}, \doi{10.1016/j.jss.2022.111251}.

\bibitemdeclare{inproceedings}{DBLP:conf/isola/AudritoDT22}
\bibitem{DBLP:conf/isola/AudritoDT22}
\bibinfo{author}{Giorgio \surnamestart Audrito\surnameend},
  \bibinfo{author}{Ferruccio \surnamestart Damiani\surnameend} \&
  \bibinfo{author}{Gianluca \surnamestart Torta\surnameend}
  (\bibinfo{year}{2022}): \emph{\bibinfo{title}{Bringing Aggregate Programming
  Towards the Cloud}}.
\newblock In: {\slshape \bibinfo{booktitle}{11th International Symposium On
  Leveraging Applications of Formal Methods, Verification and Validation, ISoLA
  2022, Proceedings, Part {III}}}, {\slshape \bibinfo{series}{LNCS}}
  \bibinfo{volume}{13703}, \bibinfo{publisher}{Springer}, pp.
  \bibinfo{pages}{301--317}, \doi{10.1007/978-3-031-19759-8_19}.

\bibitemdeclare{inproceedings}{a:rtssgradient}
\bibitem{a:rtssgradient}
\bibinfo{author}{Giorgio \surnamestart Audrito\surnameend},
  \bibinfo{author}{Ferruccio \surnamestart Damiani\surnameend},
  \bibinfo{author}{Mirko \surnamestart Viroli\surnameend} \&
  \bibinfo{author}{Enrico \surnamestart Bini\surnameend}
  (\bibinfo{year}{2018}): \emph{\bibinfo{title}{Distributed Real-Time
  Shortest-Paths Computations with the Field Calculus}}.
\newblock In: {\slshape \bibinfo{booktitle}{2018 {IEEE} Real-Time Systems
  Symposium (RTSS)}}, \bibinfo{publisher}{{IEEE} Computer Society}, pp.
  \bibinfo{pages}{23--34}, \doi{10.1109/RTSS.2018.00013}.

\bibitemdeclare{inproceedings}{DBLP:conf/coordination/AudritoRT22}
\bibitem{DBLP:conf/coordination/AudritoRT22}
\bibinfo{author}{Giorgio \surnamestart Audrito\surnameend},
  \bibinfo{author}{Luigi \surnamestart Rapetta\surnameend} \&
  \bibinfo{author}{Gianluca \surnamestart Torta\surnameend}
  (\bibinfo{year}{2022}): \emph{\bibinfo{title}{Extensible 3D Simulation of
  Aggregated Systems with {FCPP}}}.
\newblock In: {\slshape \bibinfo{booktitle}{Coordination Models and Languages -
  24th International Conference, {COORDINATION} 2022 Proceedings}}, {\slshape
  \bibinfo{series}{LNCS}} \bibinfo{volume}{13271},
  \bibinfo{publisher}{Springer}, pp. \bibinfo{pages}{55--71},
  \doi{10.1007/978-3-031-08143-9_4}.

\bibitemdeclare{article}{DBLP:journals/tpds/AudritoTF23}
\bibitem{DBLP:journals/tpds/AudritoTF23}
\bibinfo{author}{Giorgio \surnamestart Audrito\surnameend},
  \bibinfo{author}{Federico \surnamestart Terraneo\surnameend} \&
  \bibinfo{author}{William \surnamestart Fornaciari\surnameend}
  (\bibinfo{year}{2023}): \emph{\bibinfo{title}{{FCPP+Miosix}: Scaling
  Aggregate Programming to Embedded Systems}}.
\newblock {\slshape \bibinfo{journal}{{IEEE} Trans. Parallel Distributed
  Syst.}} \bibinfo{volume}{34}(\bibinfo{number}{3}), pp.
  \bibinfo{pages}{869--880}, \doi{10.1109/TPDS.2022.3232633}.

\bibitemdeclare{article}{DBLP:journals/scp/AudritoT24}
\bibitem{DBLP:journals/scp/AudritoT24}
\bibinfo{author}{Giorgio \surnamestart Audrito\surnameend} \&
  \bibinfo{author}{Gianluca \surnamestart Torta\surnameend}
  (\bibinfo{year}{2024}): \emph{\bibinfo{title}{{FCPP} to aggregate them all}}.
\newblock {\slshape \bibinfo{journal}{Sci. Comput. Program.}}
  \bibinfo{volume}{231}, p. \bibinfo{pages}{103026},
  \doi{10.1016/J.SCICO.2023.103026}.

\bibitemdeclare{article}{Viroli:HFC-TOCL}
\bibitem{Viroli:HFC-TOCL}
\bibinfo{author}{Giorgio \surnamestart Audrito\surnameend},
  \bibinfo{author}{Mirko \surnamestart Viroli\surnameend},
  \bibinfo{author}{Ferruccio \surnamestart Damiani\surnameend},
  \bibinfo{author}{Danilo \surnamestart Pianini\surnameend} \&
  \bibinfo{author}{Jacob \surnamestart Beal\surnameend} (\bibinfo{year}{2019}):
  \emph{\bibinfo{title}{A Higher-Order Calculus of Computational Fields}}.
\newblock {\slshape \bibinfo{journal}{ACM Trans. Comput. Logic}}
  \bibinfo{volume}{20}(\bibinfo{number}{1}), pp. \bibinfo{pages}{5:1--5:55},
  \doi{10.1145/3285956}.

\bibitemdeclare{incollection}{SpatialIGI2013}
\bibitem{SpatialIGI2013}
\bibinfo{author}{Jacob \surnamestart Beal\surnameend}, \bibinfo{author}{Stefan
  \surnamestart Dulman\surnameend}, \bibinfo{author}{Kyle \surnamestart
  Usbeck\surnameend}, \bibinfo{author}{Mirko \surnamestart Viroli\surnameend}
  \& \bibinfo{author}{Nikolaus \surnamestart Correll\surnameend}
  (\bibinfo{year}{2013}): \emph{\bibinfo{title}{Organizing the Aggregate:
  Languages for Spatial Computing}}.
\newblock In: {\slshape \bibinfo{booktitle}{Formal and Practical Aspects of
  Domain-Specific Languages: Recent Developments}},
  chapter~\bibinfo{chapter}{16}, \bibinfo{publisher}{IGI Global}, pp.
  \bibinfo{pages}{436--501}, \doi{10.4018/978-1-4666-2092-6.ch016}.

\bibitemdeclare{article}{BPV-COMPUTER2015}
\bibitem{BPV-COMPUTER2015}
\bibinfo{author}{Jacob \surnamestart Beal\surnameend}, \bibinfo{author}{Danilo
  \surnamestart Pianini\surnameend} \& \bibinfo{author}{Mirko \surnamestart
  Viroli\surnameend} (\bibinfo{year}{2015}): \emph{\bibinfo{title}{Aggregate
  Programming for the {Internet of Things}}}.
\newblock {\slshape \bibinfo{journal}{IEEE Computer}}
  \bibinfo{volume}{48}(\bibinfo{number}{9}), \doi{10.1109/MC.2015.261}.

\bibitemdeclare{article}{DBLP:journals/computer/BealPV15}
\bibitem{DBLP:journals/computer/BealPV15}
\bibinfo{author}{Jacob \surnamestart Beal\surnameend}, \bibinfo{author}{Danilo
  \surnamestart Pianini\surnameend} \& \bibinfo{author}{Mirko \surnamestart
  Viroli\surnameend} (\bibinfo{year}{2015}): \emph{\bibinfo{title}{Aggregate
  Programming for the Internet of Things}}.
\newblock {\slshape \bibinfo{journal}{Computer}}
  \bibinfo{volume}{48}(\bibinfo{number}{9}), pp. \bibinfo{pages}{22--30},
  \doi{10.1109/MC.2015.261}.

\bibitemdeclare{article}{BVPD-TAAS2017}
\bibitem{BVPD-TAAS2017}
\bibinfo{author}{Jacob \surnamestart Beal\surnameend}, \bibinfo{author}{Mirko
  \surnamestart Viroli\surnameend}, \bibinfo{author}{Danilo \surnamestart
  Pianini\surnameend} \& \bibinfo{author}{Ferruccio \surnamestart
  Damiani\surnameend} (\bibinfo{year}{2017}):
  \emph{\bibinfo{title}{Self-Adaptation to Device Distribution in the {Internet
  of Things}}}.
\newblock {\slshape \bibinfo{journal}{{ACM} Trans. Auton. Adapt. Syst.}}
  \bibinfo{volume}{12}(\bibinfo{number}{3}), pp. \bibinfo{pages}{12:1--12:29},
  \doi{10.1145/3105758}.

\bibitemdeclare{inproceedings}{DBLP:conf/saso/CabriH14}
\bibitem{DBLP:conf/saso/CabriH14}
\bibinfo{author}{Giacomo \surnamestart Cabri\surnameend} \&
  \bibinfo{author}{Emma \surnamestart Hart\surnameend} (\bibinfo{year}{2014}):
  \emph{\bibinfo{title}{2nd {FOCAS} Workshop on Fundamentals of Collective
  Adaptive Systems}}.
\newblock In: {\slshape \bibinfo{booktitle}{Eighth {IEEE} International
  Conference on Self-Adaptive and Self-Organizing Systems Workshops, {SASOW}
  2014, London, United Kingdom, September 8-12, 2014}},
  \bibinfo{publisher}{{IEEE} Computer Society}, pp. \bibinfo{pages}{6--7},
  \doi{10.1109/SASOW.2014.8}.

\bibitemdeclare{article}{DBLP:journals/csur/Casadei23}
\bibitem{DBLP:journals/csur/Casadei23}
\bibinfo{author}{Roberto \surnamestart Casadei\surnameend}
  (\bibinfo{year}{2023}): \emph{\bibinfo{title}{Macroprogramming: Concepts,
  State of the Art, and Opportunities of Macroscopic Behaviour Modelling}}.
\newblock {\slshape \bibinfo{journal}{{ACM} Comput. Surv.}}
  \bibinfo{volume}{55}(\bibinfo{number}{13s}), pp.
  \bibinfo{pages}{275:1--275:37}, \doi{10.1145/3579353}.

\bibitemdeclare{article}{caadpstv:tosem25}
\bibitem{caadpstv:tosem25}
\bibinfo{author}{Roberto \surnamestart Casadei\surnameend},
  \bibinfo{author}{Gianluca \surnamestart Aguzzi\surnameend},
  \bibinfo{author}{Giorgio \surnamestart Audrito\surnameend},
  \bibinfo{author}{Ferruccio \surnamestart Damiani\surnameend},
  \bibinfo{author}{Danilo \surnamestart Pianini\surnameend},
  \bibinfo{author}{Giordano \surnamestart Scarso\surnameend},
  \bibinfo{author}{Gianluca \surnamestart Torta\surnameend} \&
  \bibinfo{author}{Mirko \surnamestart Viroli\surnameend}
  (\bibinfo{year}{2025}): \emph{\bibinfo{title}{Software Engineering for
  Collective Cyber-Physical Ecosystems}}.
\newblock {\slshape \bibinfo{journal}{ACM Trans. Softw. Eng. Methodol.}}
  \bibinfo{volume}{34}(\bibinfo{number}{5}), \doi{10.1145/3712004}.

\bibitemdeclare{inproceedings}{DBLP:conf/acsos/CasadeiAPV23}
\bibitem{DBLP:conf/acsos/CasadeiAPV23}
\bibinfo{author}{Roberto \surnamestart Casadei\surnameend},
  \bibinfo{author}{Gianluca \surnamestart Aguzzi\surnameend},
  \bibinfo{author}{Danilo \surnamestart Pianini\surnameend} \&
  \bibinfo{author}{Mirko \surnamestart Viroli\surnameend}
  (\bibinfo{year}{2023}): \emph{\bibinfo{title}{Programming (and Learning)
  Self-Adaptive {\&} Self-Organising Behaviour with ScaFi: for Swarms,
  Edge-Cloud Ecosystems, and More}}.
\newblock In: {\slshape \bibinfo{booktitle}{{IEEE} International Conference on
  Autonomic Computing and Self-Organizing Systems, {ACSOS} 2023 - Companion,
  Toronto, ON, Canada, September 25-29, 2023}}, \bibinfo{publisher}{{IEEE}},
  pp. \bibinfo{pages}{33--34}, \doi{10.1109/ACSOS-C58168.2023.00032}.

\bibitemdeclare{inproceedings}{Casadei:PMLDC16}
\bibitem{Casadei:PMLDC16}
\bibinfo{author}{Roberto \surnamestart Casadei\surnameend} \&
  \bibinfo{author}{Mirko \surnamestart Viroli\surnameend}
  (\bibinfo{year}{2016}): \emph{\bibinfo{title}{Towards Aggregate Programming
  in {Scala}}}.
\newblock In: {\slshape \bibinfo{booktitle}{First Workshop on Programming
  Models and Languages for Distributed Computing}}, \bibinfo{series}{PMLDC
  '16}, \bibinfo{publisher}{ACM}, \bibinfo{address}{New York, NY, USA}, pp.
  \bibinfo{pages}{5:1--5:7}, \doi{10.1145/2957319.2957372}.

\bibitemdeclare{article}{DBLP:journals/softx/CasadeiVAP22}
\bibitem{DBLP:journals/softx/CasadeiVAP22}
\bibinfo{author}{Roberto \surnamestart Casadei\surnameend},
  \bibinfo{author}{Mirko \surnamestart Viroli\surnameend},
  \bibinfo{author}{Gianluca \surnamestart Aguzzi\surnameend} \&
  \bibinfo{author}{Danilo \surnamestart Pianini\surnameend}
  (\bibinfo{year}{2022}): \emph{\bibinfo{title}{ScaFi: {A} Scala {DSL} and
  Toolkit for Aggregate Programming}}.
\newblock {\slshape \bibinfo{journal}{SoftwareX}} \bibinfo{volume}{20}, p.
  \bibinfo{pages}{101248}, \doi{10.1016/j.softx.2022.101248}.

\bibitemdeclare{inproceedings}{Damas-Milner:POPL-1982}
\bibitem{Damas-Milner:POPL-1982}
\bibinfo{author}{Luis \surnamestart Damas\surnameend} \& \bibinfo{author}{Robin
  \surnamestart Milner\surnameend} (\bibinfo{year}{1982}):
  \emph{\bibinfo{title}{Principal Type-schemes for Functional Programs}}.
\newblock In: {\slshape \bibinfo{booktitle}{Symposium on Principles of
  Programming Languages}}, \bibinfo{series}{POPL '82},
  \bibinfo{publisher}{ACM}, pp. \bibinfo{pages}{207--212},
  \doi{10.1145/582153.582176}.

\bibitemdeclare{article}{DVB-SCP2016}
\bibitem{DVB-SCP2016}
\bibinfo{author}{Ferruccio \surnamestart Damiani\surnameend},
  \bibinfo{author}{Mirko \surnamestart Viroli\surnameend} \&
  \bibinfo{author}{Jacob \surnamestart Beal\surnameend} (\bibinfo{year}{2016}):
  \emph{\bibinfo{title}{A type-sound calculus of computational fields}}.
\newblock {\slshape \bibinfo{journal}{Science of Computer Programming}}
  \bibinfo{volume}{117}, pp. \bibinfo{pages}{17 -- 44},
  \doi{10.1016/j.scico.2015.11.005}.

\bibitemdeclare{inproceedings}{DasguptaCDC16}
\bibitem{DasguptaCDC16}
\bibinfo{author}{Soura \surnamestart Dasgupta\surnameend} \&
  \bibinfo{author}{Jacob \surnamestart Beal\surnameend} (\bibinfo{year}{2016}):
  \emph{\bibinfo{title}{A {Lyapunov} analysis for the robust stability of an
  adaptive {Bellman-Ford} algorithm}}.
\newblock In: {\slshape \bibinfo{booktitle}{Decision and Control (CDC), 2016
  IEEE 55th Conference on}}, \bibinfo{organization}{IEEE}, pp.
  \bibinfo{pages}{7282--7287}, \doi{10.1109/CDC.2016.7799393}.

\bibitemdeclare{article}{autonomicommunications}
\bibitem{autonomicommunications}
\bibinfo{author}{Simon \surnamestart Dobson\surnameend},
  \bibinfo{author}{Spyros \surnamestart Denazis\surnameend},
  \bibinfo{author}{Antonio \surnamestart Fern{\'a}ndez\surnameend},
  \bibinfo{author}{Dominique \surnamestart Ga\"{\i}ti\surnameend},
  \bibinfo{author}{Erol \surnamestart Gelenbe\surnameend},
  \bibinfo{author}{Fabio \surnamestart Massacci\surnameend},
  \bibinfo{author}{Paddy \surnamestart Nixon\surnameend},
  \bibinfo{author}{Fabrice \surnamestart Saffre\surnameend},
  \bibinfo{author}{Nikita \surnamestart Schmidt\surnameend} \&
  \bibinfo{author}{Franco \surnamestart Zambonelli\surnameend}
  (\bibinfo{year}{2006}): \emph{\bibinfo{title}{A survey of autonomic
  communications}}.
\newblock {\slshape \bibinfo{journal}{TAAS}}
  \bibinfo{volume}{1}(\bibinfo{number}{2}), pp. \bibinfo{pages}{223--259},
  \doi{10.1145/1186778.1186782}.

\bibitemdeclare{inproceedings}{kairos}
\bibitem{kairos}
\bibinfo{author}{Ramakrishna \surnamestart Gummadi\surnameend},
  \bibinfo{author}{Omprakash \surnamestart Gnawali\surnameend} \&
  \bibinfo{author}{Ramesh \surnamestart Govindan\surnameend}
  (\bibinfo{year}{2005}): \emph{\bibinfo{title}{Macro-programming Wireless
  Sensor Networks Using {Kairos}}}.
\newblock In: {\slshape \bibinfo{booktitle}{Distributed Computing in Sensor
  Systems (DCOSS)}}, pp. \bibinfo{pages}{126--140}, \doi{10.1007/11502593_12}.

\bibitemdeclare{article}{FJ}
\bibitem{FJ}
\bibinfo{author}{Atsushi \surnamestart Igarashi\surnameend},
  \bibinfo{author}{Benjamin~C. \surnamestart Pierce\surnameend} \&
  \bibinfo{author}{Philip \surnamestart Wadler\surnameend}
  (\bibinfo{year}{2001}): \emph{\bibinfo{title}{{Featherweight Java}: A Minimal
  Core Calculus for {Java} and {GJ}}}.
\newblock {\slshape \bibinfo{journal}{{ACM} Trans. Program. Lang. Syst.}}
  \bibinfo{volume}{23}(\bibinfo{number}{3}), \doi{10.1145/503502.503505}.

\bibitemdeclare{article}{DBLP:journals/corr/abs-1108-5643}
\bibitem{DBLP:journals/corr/abs-1108-5643}
\bibinfo{author}{Serge \surnamestart Kernbach\surnameend},
  \bibinfo{author}{Thomas \surnamestart Schmickl\surnameend} \&
  \bibinfo{author}{Jon \surnamestart Timmis\surnameend} (\bibinfo{year}{2011}):
  \emph{\bibinfo{title}{Collective Adaptive Systems: Challenges Beyond
  Evolvability}}.
\newblock {\slshape \bibinfo{journal}{CoRR}} \bibinfo{volume}{abs/1108.5643}.
\newblock \eprint{1108.5643}.

\bibitemdeclare{article}{lamport:events}
\bibitem{lamport:events}
\bibinfo{author}{Leslie \surnamestart Lamport\surnameend}
  (\bibinfo{year}{1978}): \emph{\bibinfo{title}{Time, Clocks, and the Ordering
  of Events in a Distributed System}}.
\newblock {\slshape \bibinfo{journal}{Commun. {ACM}}}
  \bibinfo{volume}{21}(\bibinfo{number}{7}), pp. \bibinfo{pages}{558--565},
  \doi{10.1145/359545.359563}.

\bibitemdeclare{article}{DBLP:journals/corr/Lluch-LafuenteL16}
\bibitem{DBLP:journals/corr/Lluch-LafuenteL16}
\bibinfo{author}{Alberto \surnamestart Lluch{-}Lafuente\surnameend},
  \bibinfo{author}{Michele \surnamestart Loreti\surnameend} \&
  \bibinfo{author}{Ugo \surnamestart Montanari\surnameend}
  (\bibinfo{year}{2017}): \emph{\bibinfo{title}{Asynchronous Distributed
  Execution Of Fixpoint-Based Computational Fields}}.
\newblock {\slshape \bibinfo{journal}{Log. Methods Comput. Sci.}}
  \bibinfo{volume}{13}(\bibinfo{number}{1}), \doi{10.23638/LMCS-13(1:13)2017}.

\bibitemdeclare{inproceedings}{DBLP:conf/sfm/LoretiH16}
\bibitem{DBLP:conf/sfm/LoretiH16}
\bibinfo{author}{Michele \surnamestart Loreti\surnameend} \&
  \bibinfo{author}{Jane \surnamestart Hillston\surnameend}
  (\bibinfo{year}{2016}): \emph{\bibinfo{title}{Modelling and Analysis of
  Collective Adaptive Systems with {CARMA} and its Tools}}.
\newblock In: {\slshape \bibinfo{booktitle}{Formal Methods for the Quantitative
  Evaluation of Collective Adaptive Systems - 16th International School on
  Formal Methods for the Design of Computer, Communication, and Software
  Systems, {SFM} 2016, Advanced Lectures}}, {\slshape \bibinfo{series}{LNCS}}
  \bibinfo{volume}{9700}, \bibinfo{publisher}{Springer}, pp.
  \bibinfo{pages}{83--119}, \doi{10.1007/978-3-319-34096-8_4}.

\bibitemdeclare{article}{DBLP:journals/jsa/MameiMTZ06}
\bibitem{DBLP:journals/jsa/MameiMTZ06}
\bibinfo{author}{Marco \surnamestart Mamei\surnameend},
  \bibinfo{author}{Ronaldo \surnamestart Menezes\surnameend},
  \bibinfo{author}{Robert \surnamestart Tolksdorf\surnameend} \&
  \bibinfo{author}{Franco \surnamestart Zambonelli\surnameend}
  (\bibinfo{year}{2006}): \emph{\bibinfo{title}{Case studies for
  self-organization in computer science}}.
\newblock {\slshape \bibinfo{journal}{J. Syst. Archit.}}
  \bibinfo{volume}{52}(\bibinfo{number}{8-9}), pp. \bibinfo{pages}{443--460},
  \doi{10.1016/j.sysarc.2006.02.002}.

\bibitemdeclare{inproceedings}{Mamei:2004a}
\bibitem{Mamei:2004a}
\bibinfo{author}{Marco \surnamestart Mamei\surnameend} \&
  \bibinfo{author}{Franco \surnamestart Zambonelli\surnameend}
  (\bibinfo{year}{2004}): \emph{\bibinfo{title}{Programming pervasive and
  mobile computing applications with the {TOTA} middleware}}.
\newblock In: {\slshape \bibinfo{booktitle}{Pervasive Computing and
  Communications, 2004}}, \bibinfo{publisher}{IEEE}, pp. \bibinfo{pages}{263--
  273}, \doi{10.1109/PERCOM.2004.1276864}.

\bibitemdeclare{inproceedings}{DBLP:conf/esaw/MameiZL02}
\bibitem{DBLP:conf/esaw/MameiZL02}
\bibinfo{author}{Marco \surnamestart Mamei\surnameend}, \bibinfo{author}{Franco
  \surnamestart Zambonelli\surnameend} \& \bibinfo{author}{Letizia
  \surnamestart Leonardi\surnameend} (\bibinfo{year}{2002}):
  \emph{\bibinfo{title}{Co-Fields: Towards a Unifying Approach to the
  Engineering of Swarm Intelligent Systems}}.
\newblock In: {\slshape \bibinfo{booktitle}{3rd International Workshop on
  Engineering Societies in the Agents World}}, {\slshape
  \bibinfo{series}{LNCS}} \bibinfo{volume}{2577},
  \bibinfo{publisher}{Springer}, pp. \bibinfo{pages}{68--81},
  \doi{10.1007/3-540-39173-8\_6}.

\bibitemdeclare{inproceedings}{regiment}
\bibitem{regiment}
\bibinfo{author}{Ryan \surnamestart Newton\surnameend} \& \bibinfo{author}{Matt
  \surnamestart Welsh\surnameend} (\bibinfo{year}{2004}):
  \emph{\bibinfo{title}{Region Streams: Functional Macroprogramming for Sensor
  Networks}}.
\newblock In: {\slshape \bibinfo{booktitle}{Workshop on Data Management for
  Sensor Networks}}, pp. \bibinfo{pages}{78--87},
  \doi{10.1145/1052199.1052213}.

\bibitemdeclare{article}{DBLP:journals/scp/NicolaDIT18}
\bibitem{DBLP:journals/scp/NicolaDIT18}
\bibinfo{author}{Rocco~De \surnamestart Nicola\surnameend},
  \bibinfo{author}{Tan \surnamestart Duong\surnameend}, \bibinfo{author}{Omar
  \surnamestart Inverso\surnameend} \& \bibinfo{author}{Catia \surnamestart
  Trubiani\surnameend} (\bibinfo{year}{2018}): \emph{\bibinfo{title}{AErlang:
  Empowering Erlang with attribute-based communication}}.
\newblock {\slshape \bibinfo{journal}{Sci. Comput. Program.}}
  \bibinfo{volume}{168}, pp. \bibinfo{pages}{71--93},
  \doi{10.1016/j.scico.2018.08.006}.

\bibitemdeclare{article}{DBLP:journals/sttt/NicolaJW20}
\bibitem{DBLP:journals/sttt/NicolaJW20}
\bibinfo{author}{Rocco~De \surnamestart Nicola\surnameend},
  \bibinfo{author}{Stefan \surnamestart J{\"{a}}hnichen\surnameend} \&
  \bibinfo{author}{Martin \surnamestart Wirsing\surnameend}
  (\bibinfo{year}{2020}): \emph{\bibinfo{title}{Rigorous engineering of
  collective adaptive systems: special section}}.
\newblock {\slshape \bibinfo{journal}{Int. J. Softw. Tools Technol. Transf.}}
  \bibinfo{volume}{22}(\bibinfo{number}{4}), pp. \bibinfo{pages}{389--397},
  \doi{10.1007/s10009-020-00565-0}.

\bibitemdeclare{article}{DBLP:journals/taas/NicolaLPT14}
\bibitem{DBLP:journals/taas/NicolaLPT14}
\bibinfo{author}{Rocco~De \surnamestart Nicola\surnameend},
  \bibinfo{author}{Michele \surnamestart Loreti\surnameend},
  \bibinfo{author}{Rosario \surnamestart Pugliese\surnameend} \&
  \bibinfo{author}{Francesco \surnamestart Tiezzi\surnameend}
  (\bibinfo{year}{2014}): \emph{\bibinfo{title}{A Formal Approach to Autonomic
  Systems Programming: The {SCEL} Language}}.
\newblock {\slshape \bibinfo{journal}{{ACM} Trans. Auton. Adapt. Syst.}}
  \bibinfo{volume}{9}(\bibinfo{number}{2}), pp. \bibinfo{pages}{7:1--7:29},
  \doi{10.1145/2619998}.

\bibitemdeclare{incollection}{nitti2025drone}
\bibitem{nitti2025drone}
\bibinfo{author}{L~\surnamestart Nitti\surnameend},
  \bibinfo{author}{D~\surnamestart Bortoluzzi\surnameend},
  \bibinfo{author}{A~\surnamestart Biglia\surnameend},
  \bibinfo{author}{D~Ricauda \surnamestart Aimonino\surnameend},
  \bibinfo{author}{G~\surnamestart Torta\surnameend},
  \bibinfo{author}{G~\surnamestart Audrito\surnameend},
  \bibinfo{author}{F~\surnamestart Damiani\surnameend},
  \bibinfo{author}{P~\surnamestart Gay\surnameend},
  \bibinfo{author}{A~\surnamestart Rapp\surnameend} \&
  \bibinfo{author}{L~\surnamestart Comba\surnameend} (\bibinfo{year}{2025}):
  \emph{\bibinfo{title}{Drone-swarm based surveillance system for autonomous
  machine safety functionality}}.
\newblock In: {\slshape \bibinfo{booktitle}{Precision agriculture'25}},
  \bibinfo{publisher}{Wageningen Academic}, pp. \bibinfo{pages}{592--598},
  \doi{10.1163/9789004725232_077}.

\bibitemdeclare{article}{DBLP:journals/software/PinciroliB16}
\bibitem{DBLP:journals/software/PinciroliB16}
\bibinfo{author}{Carlo \surnamestart Pinciroli\surnameend} \&
  \bibinfo{author}{Giovanni \surnamestart Beltrame\surnameend}
  (\bibinfo{year}{2016}): \emph{\bibinfo{title}{Buzz: {A} Programming Language
  for Robot Swarms}}.
\newblock {\slshape \bibinfo{journal}{{IEEE} Softw.}}
  \bibinfo{volume}{33}(\bibinfo{number}{4}), pp. \bibinfo{pages}{97--100},
  \doi{10.1109/MS.2016.95}.

\bibitemdeclare{article}{DBLP:journals/percom/TestaADT22}
\bibitem{DBLP:journals/percom/TestaADT22}
\bibinfo{author}{Lorenzo \surnamestart Testa\surnameend},
  \bibinfo{author}{Giorgio \surnamestart Audrito\surnameend},
  \bibinfo{author}{Ferruccio \surnamestart Damiani\surnameend} \&
  \bibinfo{author}{Gianluca \surnamestart Torta\surnameend}
  (\bibinfo{year}{2022}): \emph{\bibinfo{title}{Aggregate processes as
  distributed adaptive services for the Industrial Internet of Things}}.
\newblock {\slshape \bibinfo{journal}{Pervasive Mob. Comput.}}
  \bibinfo{volume}{85}, p. \bibinfo{pages}{101658},
  \doi{10.1016/j.pmcj.2022.101658}.

\bibitemdeclare{article}{Viroli:TOMACSselfstab}
\bibitem{Viroli:TOMACSselfstab}
\bibinfo{author}{Mirko \surnamestart Viroli\surnameend},
  \bibinfo{author}{Giorgio \surnamestart Audrito\surnameend},
  \bibinfo{author}{Jacob \surnamestart Beal\surnameend},
  \bibinfo{author}{Ferruccio \surnamestart Damiani\surnameend} \&
  \bibinfo{author}{Danilo \surnamestart Pianini\surnameend}
  (\bibinfo{year}{2018}): \emph{\bibinfo{title}{Engineering Resilient
  Collective Adaptive Systems by Self-Stabilisation}}.
\newblock {\slshape \bibinfo{journal}{ACM Trans. Model. Comput. Simul.}}
  \bibinfo{volume}{28}(\bibinfo{number}{2}), pp. \bibinfo{pages}{16:1--16:28},
  \doi{10.1145/3177774}.

\bibitemdeclare{article}{DBLP:journals/jlap/ViroliBDACP19}
\bibitem{DBLP:journals/jlap/ViroliBDACP19}
\bibinfo{author}{Mirko \surnamestart Viroli\surnameend}, \bibinfo{author}{Jacob
  \surnamestart Beal\surnameend}, \bibinfo{author}{Ferruccio \surnamestart
  Damiani\surnameend}, \bibinfo{author}{Giorgio \surnamestart
  Audrito\surnameend}, \bibinfo{author}{Roberto \surnamestart
  Casadei\surnameend} \& \bibinfo{author}{Danilo \surnamestart
  Pianini\surnameend} (\bibinfo{year}{2019}): \emph{\bibinfo{title}{From
  distributed coordination to field calculus and aggregate computing}}.
\newblock {\slshape \bibinfo{journal}{J. Log. Algebraic Methods Program.}}
  \bibinfo{volume}{109}, \doi{10.1016/j.jlamp.2019.100486}.

\bibitemdeclare{inbook}{wadaa2010chronus}
\bibitem{wadaa2010chronus}
\bibinfo{author}{Hiroshi \surnamestart Wadaa\surnameend},
  \bibinfo{author}{Pruet \surnamestart Boonmab\surnameend} \&
  \bibinfo{author}{Junichi \surnamestart Suzukic\surnameend}
  (\bibinfo{year}{2010}): \emph{\bibinfo{title}{Chronus: A spatiotemporal
  macroprogramming language for autonomic wireless sensor networks}}.
\newblock \bibinfo{volume}{167}, \bibinfo{publisher}{Academic Press},
  \doi{10.1016/B978-0-12-382190-4.00008-5}.

\bibitemdeclare{article}{wang2024survey}
\bibitem{wang2024survey}
\bibinfo{author}{Lei \surnamestart Wang\surnameend}, \bibinfo{author}{Chen
  \surnamestart Ma\surnameend}, \bibinfo{author}{Xueyang \surnamestart
  Feng\surnameend}, \bibinfo{author}{Zeyu \surnamestart Zhang\surnameend},
  \bibinfo{author}{Hao \surnamestart Yang\surnameend}, \bibinfo{author}{Jingsen
  \surnamestart Zhang\surnameend}, \bibinfo{author}{Zhiyuan \surnamestart
  Chen\surnameend}, \bibinfo{author}{Jiakai \surnamestart Tang\surnameend},
  \bibinfo{author}{Xu~\surnamestart Chen\surnameend}, \bibinfo{author}{Yankai
  \surnamestart Lin\surnameend} et~al. (\bibinfo{year}{2024}):
  \emph{\bibinfo{title}{A survey on large language model based autonomous
  agents}}.
\newblock {\slshape \bibinfo{journal}{Frontiers of Computer Science}}
  \bibinfo{volume}{18}(\bibinfo{number}{6}), p. \bibinfo{pages}{186345},
  \doi{10.1007/s11704-024-40231-1}.

\bibitemdeclare{inproceedings}{DBLP:conf/fmco/WirsingHTZ11}
\bibitem{DBLP:conf/fmco/WirsingHTZ11}
\bibinfo{author}{Martin \surnamestart Wirsing\surnameend},
  \bibinfo{author}{Matthias~M. \surnamestart H{\"{o}}lzl\surnameend},
  \bibinfo{author}{Mirco \surnamestart Tribastone\surnameend} \&
  \bibinfo{author}{Franco \surnamestart Zambonelli\surnameend}
  (\bibinfo{year}{2011}): \emph{\bibinfo{title}{{ASCENS:} Engineering Autonomic
  Service-Component Ensembles}}.
\newblock In: {\slshape \bibinfo{booktitle}{Formal Methods for Components and
  Objects, 10th International Symposium, {FMCO} 2011, Revised Selected
  Papers}}, {\slshape \bibinfo{series}{LNCS}} \bibinfo{volume}{7542},
  \bibinfo{publisher}{Springer}, pp. \bibinfo{pages}{1--24},
  \doi{10.1007/978-3-642-35887-6_1}.

\end{thebibliography}

\end{document}